\documentclass[fleqn,usenatbib]{mnras}

\usepackage{newtxtext,newtxmath}

\usepackage[T1]{fontenc}

\DeclareRobustCommand{\VAN}[3]{#2}
\let\VANthebibliography\thebibliography
\def\thebibliography{\DeclareRobustCommand{\VAN}[3]{##3}\VANthebibliography}

\usepackage{graphicx}	
\usepackage{amsmath,bm}	
\usepackage{pdflscape}
\usepackage{color, soul}
\usepackage{comment}

\graphicspath{{Figures/}}
\newcommand{\feh}{\left[\frac{Fe}{H}\right]}

\title[Tidal Circularization of Exoplanet Systems]{Measuring Tidal Dissipation in Giant Planets from Tidal Circularization}

\author[M. Mahmud et al.]{
    Mohammad M. Mahmud,$^{1}$
    Kaloyan M. Penev,$^{1}$\thanks{E-mail: kaloyan.penev@utdallas.edu (KMP)}
    Joshua A. Schussler$^{1}$
\\
$^{1}$Department of Physics, University of Texas at Dallas, 800 W. Campbell Rd.,
Richardson, TX 75080, USA\\
}

\date{Accepted XXX. Received YYY; in original form ZZZ}

\pubyear{2023}

\begin{document}
\label{firstpage}
\pagerange{\pageref{firstpage}--\pageref{lastpage}}
\maketitle

\begin{abstract}
    In this project, we determined the constraints on the modified tidal quality
    factor, $Q_{pl}'$, of gas-giant planets orbiting close to their host stars.
    We allowed $Q_{pl}'$ to depend on tidal frequency, accounting for the
    multiple tidal waves with time-dependent frequencies simultaneously present
    on the planet.  We performed our analysis on 78 single-star and
    single-planet systems, with giant planets and host stars with radiative
    cores and convective outer shells. We extracted constraints on the
    frequency-dependent $Q_{pl}'$ for each system separately and combined them
    to find general constraints on $Q_{pl}'$ required to explain the observed
    eccentricity envelope while simultaneously allowing the observed
    eccentricities of all systems to survive to the present day. Individual
        systems do not place tight constraints on $Q_{pl}'$. However, since
        similar planets must have similar tidal dissipation, we require that a
    consistent, possibly frequency-dependent, model must apply. Under that
assumption, we find that the value of $\log_{10}Q_{pl}'$ for HJs is $5.0\pm0.5$
for the range of tidal period from 0.8 to 7 days. We did not see any clear sign
of frequency dependence of $Q_{pl}'$.
\end{abstract}




\section{Introduction}\label{introduction}

In a star-planet system, both the star and the planet revolve around the
barycenter, in generally elliptic orbits. The changing distance between the
objects and the variable orbital speed lead to time-varying tidal deformations
of both the planet and the star. These deformations are subject to tidal
friction, converting mechanical energy to heat. As the system loses energy, the
eccentricity of the orbit reduces with time. Finally, either the planet is
engulfed by the star or the planet's orbit becomes circular. This process is
called the tidal circularization of the planet's orbit.

This article focuses on the tidal dissipation within hot Jupiters (HJs): gas giant
exoplanets observed in orbits with periods shorter than 10 days. These planets
pose a challenge to our understanding of planet formation. Traditional planet
formation theory predicts such massive planets can only form a long distance
away from the parent star.

Several theories have been proposed to explain the existence of HJs.
For example, High Eccentricity Migration theory suggests that initially
the gas giant is formed at a long distance away from the parent star. Then,
due to some interactions like Kozai-Lidov oscillation and/or planet-planet
scattering, the eccentricity of the planet is excited to extreme values,
placing the pericenter close to the star. The short planet-star separation
at the pericenter allows tidal circularization to occur, eventually producing a
hot Jupiter \citep{Rasio_Ford_96, Fabrycky_Tremaine_07, Beauge_Nesvorny_12,
Vick_Lai_Anderson_19, hamers2017hot}.

Tides are important for other hot Jupiter formation theories as well. Two additional
favored scenarios suggest planets may migrate through interactions with the
protoplanetary disk \citep[see][for a relatively recent
review]{Baruteau_et_al_14}, or even form in situ \citep{Batygin_et_al_16,
Boley_et_al_16}. Naively, one would expect disk migration and in situ formation
to produce planets on circular orbits. However, observations belie that
expectation, since HJs far enough from their stars to not suffer tidal
circularization exhibit a broad range of eccentricities. Regardless of how HJ
eccentricities come to be excited, understanding tidal dissipation is required
to work backward in time to find the orbital configurations that disk migration
or in situ formation must explain if they are to be the dominant channel of hot
Jupiter formation. Hence, studying the tidal dissipation efficiency of the
star-planet systems is important for deciphering the mystery of hot Jupiters.

Moreover, tidal efficiency can give us an idea about the internal structure of
a planet. Rocky planets have solid layers in their core, so due to tidal
distortion, their frictional energy loss is higher than that of gas giants.
Therefore, usually, rocky planets have higher tidal efficiency than gas
giants. In the solar system, the tidal dissipation of the rocky planets is 3
orders of magnitude more efficient than that of the gas giants
\citep{goldreich1966q}.

In this article, we will describe tidal dissipation using the tidal quality
factor ($Q$). This is the most widely used parameterization of tidal dissipation
in the literature. Formally, $1/Q$ is the fraction of the energy in a tidal wave
lost for each radian the wave travels. In the orbital evolution equations, tidal
dissipation always appears in combination with $k$, the tidal Love number (the
ratio of the quadrupole potential an object generates to the quadrupole tidal
potential it experiences). For that reason, we will not use $Q$ directly, but
rather $Q' \equiv Q/k$. We stress that $Q'$ is inversely related to the
dissipation, i.e. small $Q'$ corresponds to large dissipation and vice-versa.
We will denote by $Q'_{pl}$ the tidal quality factor of the planet and by
$Q'_{\star}$ that of the star. We have focused on measuring $Q'_{pl}$, by
studying tidal circularization, where the dissipation in the planet dominates.
However, the dissipation in the star also plays a role, though generally
assumed to be sub-dominant, which we must account for.

A number of theoretical models have been proposed for tidal dissipation.
Some models suggest HJ dissipation is dominated by visco-elastic deformations in
an assumed solid core \citep[e.g.][]{Remus_et_al_15, Storch_Lai_15,
Shoji_Hussmann_17}. Others expect the turbulent dissipation of tidally excited
inertial waves in the gas envelope to dominate \citep[e.g.][]{Ogilvie_13,
Ogilvie_14}. Yet other authors point out that the inertial wave dissipation
models may be missing important physics, changing the dissipation dramatically
\citep{Mathis_et_al_16, Pontin_et_al_20, Lin_21}. \citet{Guenel_et_al_14} argue
for a comparable contribution from the core and the inertial wave dissipation in
the envelope, \citet{Auclair-Desrotour_Leconte_18} propose that thermal tides
may prevent tidal synchronization of HJ atmospheres, and finally
\citet{Andre_et_al_17, Andre_et_al_19} propose that recent observation of
Jupiter suggests the presence of layered semi-convection throughout a large
fraction of the planet's envelope and argue theoretically that this could
enhance tidal dissipation by orders of magnitude. Given that we have limited
understanding of the deep interiors of HJs, the best option for narrowing down
the theoretical possibilities rests with deriving empirical constraints on
$Q_{pl}'$.

A number of efforts use the distribution of hot Jupiter eccentricities to
calibrate the value of $Q_{pl}'$, or some other tidal dissipation parameter.
\citet{bonomo2017gaps} required that the timescale for orbital circularization
should be shorter than the age of those systems which have already been
circularized, and longer than the age of systems observed in eccentric orbits,
finding a range of allowed $Q_{pl}'$ values for each system.
\citet{jackson2008tidal} ran tidal evolution backward from the present orbits of
hot Jupiters, arguing that for the true value of $Q_{pl}'$ and $Q_\star'$ the
initial eccentricity distribution for planets with semimajor axes below and above
0.2 AU should be statistically indistinguishable, finding $Q_{pl}'\sim10^{6.5}$.
\cite{quinn2014hd} argue that for the true value of $Q_{pl}'$ the
statistical difference between systems with age above vs below the tidal
circularization timescale should be maximized, finding $Q_{pl}'\sim10^6$.
\citet{Hansen_10, Hansen_12, Oconnor_Hansen_18} calibrated a different tidal
dissipation parameter (not equivalent to $Q'$) by requiring that tides
simultaneously produce a period-eccentricity envelope (i.e. at a given orbital
period tides should circularize systems to below some eccentricity no matter the
initial conditions) while at the same time allowing the observed eccentricity of
each system to survive to the present age of the system.  Converting their tidal
parameter to $Q_{pl}'$, these authors find a significantly higher value of
$\sim10^{7.5}$. Clearly, even though all these analyses are based on the
same tidal effect (circularization of the orbit) for the same class of objects
(hot Jupiters), there is a disagreement of more than an order of magnitude
between them.

The above analyses all treat tides through a single parameter that is assumed
constant throughout the evolution. In contrast, all proposed tidal dissipation
mechanisms predict that the dissipation will change with tidal frequency, though
different theoretical models predict different frequency dependencies. In this
paper, we present a detailed analysis of tidal circularization in hot Jupiter
systems, allowing for frequency-dependent tides. We also include the effect of
both stellar and planetary tides on the orbital evolution, the evolution of the
stellar radius and internal structure, the evolution of the spin of the star
under magnetic braking, tidal torques, and internal differential rotation.
We study each star-exoplanet system separately, using Bayesian analysis to
account for observational and model uncertainties and unknown initial
conditions. Final constraints for the dissipation are constructed by combining
the results of individual systems.

Section \ref{sec:data} describes the collection of exoplanet system parameters used
in the analysis. Section \ref{sec:evolution} describes the tidal evolution model
that we use to follow the orbital evolution. Section \ref{sec:q_prescription} defines
our model for frequency-dependent $Q_{pl}'$. Section \ref{sec:TidalCircularization}
describes the procedure we use to constrain the dissipation from the observed
distribution of hot Jupiter eccentricities. Section \ref{sec:bayesianAnalysis}
describes the details of the MCMC analysis we perform for each system to extract
individual frequency-dependent constraints on $Q_{pl}'$. Section
\ref{sec:individual_constraints} presents the individual constraints for each system along with various convergence diagnostics. Section \ref{sec:combined_constraints} shows the combined constraints for the individual systems under the assumption that a common, though frequency-dependent, dissipation applies for all planets in our sample. Section \ref{sec:comparison} places our results in context with previous empirical tidal dissipation constraints from the literature. Finally, Section 6 summarizes our findings.

\section{Input Data}
\label{sec:data}

\begin{table*}
    \centering
    \caption{
        Data on the star-exoplanet systems. The values of $M_\star$, $M_{p}$,
        $R_\star$, $R_{p}$, $a$, $P_{orb}$ and $t$ are recorded in the unit of the
        solar mass, Jupiter mass, solar radius, Jupiter radius, AU, days and
        Gyr, respectively
    } \label{tab:exotable}
    \resizebox{\textwidth}{!}{\begin{tabular}{lccccccccccl}
\hline
System & $M_{\star} (M_{\odot})$ & $M_{p} (M_{Jupiter})$ & $[Fe/H]_{\star}$ & $\frac{R_{p}^2}{R_{\star}^2}\Big(\%\Big)$ &  $R_{\star} (R_{\odot})$ & $R_{p} (R_{Jupiter})$ & $a$ (AU)  & $P_{orb}$(days)  & $t$ (Gyr) & $e$ & Reference\\
\hline
CoRoT-5 b & $1.00_{-0.02}^{+0.02}$ & $0.46700_{-0.02400}^{+0.04700}$ & $-0.250_{-0.060}^{+0.060}$ & $1.4610_{-0.0320}^{+0.0300}$ & $1.19_{-0.04}^{+0.04}$ & $1.388_{-0.047}^{+0.046}$ & $0.04947_{-0.00029}^{+0.00026}$ & $4.0378962_{-0.00000190}^{+0.00000190}$ & $6.90_{-1.40}^{+1.40}$ & $0.0900_{-0.0400}^{+0.0900}$ & \cite{rauer2009transiting} \\
CoRoT-12 b & $1.08_{-0.07}^{+0.08}$ & $0.91700_{-0.06500}^{+0.07000}$ & $0.160_{-0.100}^{+0.100}$ & $1.7440_{-0.0400}^{+0.0390}$ & $1.12_{-0.09}^{+0.10}$ & $1.440_{-0.130}^{+0.130}$ & $0.04016_{-0.00092}^{+0.00093}$ & $2.8280420_{-0.00001300}^{+0.00001300}$ & $6.30_{-3.10}^{+3.10}$ & $0.0700_{-0.0420}^{+0.0630}$ & \cite{gillon2010transiting} \\
CoRoT-16 b & $1.10_{-0.08}^{+0.08}$ & $0.53500_{-0.08300}^{+0.08500}$ & $0.190_{-0.060}^{+0.060}$ & $1.0200_{-0.0920}^{+0.0950}$ & $1.19_{-0.13}^{+0.14}$ & $1.170_{-0.140}^{+0.160}$ & $0.06180_{-0.00150}^{+0.00150}$ & $5.3522700_{-0.00020000}^{+0.00020000}$ & $6.73_{-2.80}^{+2.80}$ & $0.3300_{-0.1000}^{+0.0900}$ & \cite{ollivier2012transiting} \\
CoRoT-23 b & $1.14_{-0.08}^{+0.08}$ & $2.80000_{-0.30000}^{+0.30000}$ & $0.050_{-0.100}^{+0.100}$ & -- & $1.61_{-0.18}^{+0.18}$ & $1.050_{-0.130}^{+0.130}$ & $0.04800_{-0.00400}^{+0.00400}$ & $3.6313000_{-0.00010000}^{+0.00010000}$ & $7.20_{-1.00}^{+1.50}$ & $0.1600_{-0.0200}^{+0.0200}$ & \cite{rouan2012transiting} \\
CoRoT-27 b & $1.05_{-0.11}^{+0.11}$ & $10.39000_{-0.55000}^{+0.55000}$ & $-0.100_{-0.100}^{+0.100}$ & -- & $1.08_{-0.06}^{+0.18}$ & $1.007_{-0.044}^{+0.044}$ & $0.04760_{-0.00660}^{+0.00660}$ & $3.5753200_{-0.00006000}^{+0.00006000}$ & $4.21_{-2.72}^{+2.72}$ & e<0.065(<98 \%) & \cite{parviainen2014transiting} \\
CoRoT-29 b & $0.97_{-0.14}^{+0.14}$ & $0.85000_{-0.20000}^{+0.20000}$ & $0.200_{-0.100}^{+0.100}$ & -- & $0.90_{-0.12}^{+0.12}$ & $0.900_{-0.160}^{+0.160}$ & $0.03860_{-0.00590}^{+0.00590}$ & $2.8505700_{-0.00000600}^{+0.00000600}$ & $4.50_{-3.50}^{+3.50}$ & $0.0820_{-0.0810}^{+0.0810}$ & \cite{cabrera2015transiting} \\
CoRoT-30 b & $0.98_{-0.05}^{+0.03}$ & $2.90000_{-0.22000}^{+0.22000}$ & $0.020_{-0.100}^{+0.100}$ & -- & $0.91_{-0.03}^{+0.09}$ & $1.009_{-0.076}^{+0.076}$ & $0.08440_{-0.00120}^{+0.00120}$ & $9.0600500_{-0.00024000}^{+0.00024000}$ & $3.00_{-2.40}^{+3.70}$ & e<0.007(<84.1 \%) & \cite{borde2010transiting} \\
HAT-P-19 b & $0.84_{-0.04}^{+0.04}$ & $0.29200_{-0.01800}^{+0.01800}$ & $0.230_{-0.080}^{+0.080}$ & -- & $0.82_{-0.05}^{+0.05}$ & $1.132_{-0.072}^{+0.072}$ & $0.04660_{-0.00080}^{+0.00080}$ & $4.0087780_{-0.00000600}^{+0.00000600}$ & $8.80_{-5.20}^{+5.20}$ & $0.0670_{-0.0420}^{+0.0420}$ & \cite{hartman2010hat} \\
HAT-P-20 b & $0.76_{-0.03}^{+0.03}$ & $7.24600_{-0.18700}^{+0.18700}$ & $0.350_{-0.080}^{+0.080}$ & $0.9898$ & $0.69_{-0.02}^{+0.02}$ & $0.867_{-0.033}^{+0.033}$ & $0.03610_{-0.00050}^{+0.00050}$ & $2.8753170_{-0.00000400}^{+0.00000400}$ & $6.70_{-3.80}^{+5.70}$ & $0.0150_{-0.0050}^{+0.0050}$ & \cite{bakos2011hat} \\
HAT-P-25 b & $1.01_{-0.03}^{+0.03}$ & $0.56700_{-0.02200}^{+0.02200}$ & $0.310_{-0.080}^{+0.080}$ & $1.3811$ & $0.96_{-0.04}^{+0.05}$ & $1.190_{-0.056}^{+0.081}$ & $0.04660_{-0.00050}^{+0.00050}$ & $3.6528360_{-0.00001900}^{+0.00001900}$ & $3.20_{-2.30}^{+2.30}$ & $0.0320_{-0.0220}^{+0.0220}$ & \cite{quinn2011hat} \\
HAT-P-28 b & $1.02_{-0.05}^{+0.05}$ & $0.62600_{-0.03700}^{+0.03700}$ & $0.120_{-0.080}^{+0.080}$ & $1.4719$ & $1.10_{-0.07}^{+0.09}$ & $1.212_{-0.082}^{+0.113}$ & $0.04340_{-0.00070}^{+0.00070}$ & $3.2572150_{-0.00000700}^{+0.00000700}$ & $6.10_{-1.90}^{+2.60}$ & $0.0510_{-0.0330}^{+0.0330}$ & \cite{buchhave2011hat} \\
HAT-P-36 b & $1.02_{-0.05}^{+0.05}$ & $1.85000_{-0.10000}^{+0.10000}$ & $0.260_{-0.100}^{+0.100}$ & -- & $1.10_{-0.06}^{+0.06}$ & $1.264_{-0.071}^{+0.071}$ & $0.02382_{-0.00039}^{+0.00038}$ & $1.3273470_{-0.00000300}^{+0.00000300}$ & $6.60_{-1.80}^{+2.90}$ & e<0.059(<84.1 \%) & \cite{bonomo2017gaps} \\
HAT-P-37 b & $0.93_{-0.04}^{+0.04}$ & $1.20000_{-0.11000}^{+0.12000}$ & $0.030_{-0.100}^{+0.100}$ & -- & $0.88_{-0.04}^{+0.06}$ & $1.178_{-0.077}^{+0.077}$ & $0.03793_{-0.00058}^{+0.00056}$ & $2.7974360_{-0.00000700}^{+0.00000700}$ & $3.60_{-2.20}^{+4.10}$ & e<0.14(<84.1 \%) & \cite{bonomo2017gaps} \\
HAT-P-51 b & $0.98_{-0.03}^{+0.03}$ & $0.30900_{-0.01800}^{+0.01800}$ & $0.270_{-0.080}^{+0.080}$ & -- & $1.04_{-0.03}^{+0.04}$ & $1.293_{-0.054}^{+0.054}$ & $0.05069_{-0.00049}^{+0.00049}$ & $4.2180278_{-0.00000590}^{+0.00000590}$ & $8.20_{-1.70}^{+1.70}$ & e<0.123(<95 \%) & \cite{hartman2015hat} \\
HAT-P-52 b & $0.89_{-0.03}^{+0.03}$ & $0.81800_{-0.02900}^{+0.02900}$ & $0.280_{-0.080}^{+0.080}$ & -- & $0.89_{-0.05}^{+0.05}$ & $1.009_{-0.072}^{+0.072}$ & $0.03694_{-0.00038}^{+0.00038}$ & $2.7535953_{-0.00000940}^{+0.00000940}$ & $9.40_{-4.10}^{+4.10}$ & e<0.047(<95 \%) & \cite{hartman2015hat} \\
HAT-P-53 b & $1.09_{-0.04}^{+0.04}$ & $1.48400_{-0.05600}^{+0.05600}$ & $0.000_{-0.080}^{+0.080}$ & -- & $1.21_{-0.06}^{+0.08}$ & $1.318_{-0.091}^{+0.091}$ & $0.03159_{-0.00042}^{+0.00042}$ & $1.9616241_{-0.00000390}^{+0.00000390}$ & $4.67_{-0.83}^{+1.45}$ & e<0.134(<95 \%) & \cite{hartman2015hat} \\
HAT-P-54 b & $0.65_{-0.02}^{+0.02}$ & $0.76000_{-0.03200}^{+0.03200}$ & $-0.127_{-0.080}^{+0.080}$ & -- & $0.62_{-0.01}^{+0.01}$ & $0.944_{-0.028}^{+0.028}$ & $0.04117_{-0.00043}^{+0.00043}$ & $3.7998470_{-0.00001400}^{+0.00001400}$ & $3.90_{-2.10}^{+4.30}$ & e<0.074(<95 \%) & \cite{bakos2015hat} \\
HAT-P-55 b & $1.01_{-0.04}^{+0.04}$ & $0.58200_{-0.05600}^{+0.05600}$ & $-0.030_{-0.080}^{+0.080}$ & -- & $1.01_{-0.04}^{+0.04}$ & $1.182_{-0.055}^{+0.055}$ & $0.04604_{-0.00056}^{+0.00056}$ & $3.5852467_{-0.00000640}^{+0.00000640}$ & $4.20_{-1.70}^{+1.70}$ & e<0.139(<95 \%) & \cite{juncher2015hat} \\
HAT-P-58 b & $1.03_{-0.03}^{+0.03}$ & $0.37200_{-0.03000}^{+0.03000}$ & $-0.224_{-0.057}^{+0.057}$ & -- & $1.53_{-0.03}^{+0.03}$ & $1.332_{-0.043}^{+0.043}$ & $0.04994_{-0.00044}^{+0.00044}$ & $4.0138379_{-0.00000240}^{+0.00000240}$ & $7.11_{-0.72}^{+0.27}$ & e<0.073(<95 \%) & \cite{bakos2021hat} \\
HAT-P-59 b & $1.01_{-0.02}^{+0.02}$ & $1.54000_{-0.06700}^{+0.06700}$ & $0.217_{-0.049}^{+0.049}$ & -- & $1.10_{-0.01}^{+0.01}$ & $1.123_{-0.013}^{+0.013}$ & $0.05064_{-0.00037}^{+0.00037}$ & $4.1419771_{-0.00000120}^{+0.00000120}$ & $7.30_{-1.00}^{+1.00}$ & e<0.03(<95 \%) & \cite{bakos2021hat} \\
HAT-P-61 b & $1.00_{-0.03}^{+0.03}$ & $1.05700_{-0.07000}^{+0.07000}$ & $0.194_{-0.060}^{+0.060}$ & -- & $0.94_{-0.01}^{+0.01}$ & $0.899_{-0.027}^{+0.027}$ & $0.03010_{-0.00034}^{+0.00034}$ & $1.9023129_{-0.00000077}^{+0.00000077}$ & $2.60_{-2.00}^{+2.00}$ & e<0.113(<95 \%) & \cite{bakos2021hat} \\
HAT-P-62 b & $1.02_{-0.02}^{+0.02}$ & $0.76100_{-0.08800}^{+0.08800}$ & $0.414_{-0.090}^{+0.090}$ & -- & $1.17_{-0.02}^{+0.02}$ & $1.073_{-0.029}^{+0.029}$ & $0.03772_{-0.00024}^{+0.00024}$ & $2.6453235_{-0.00000390}^{+0.00000390}$ & $8.10_{-1.10}^{+1.10}$ & e<0.101(<95 \%) & \cite{bakos2021hat} \\
HAT-P-63 b & $0.93_{-0.02}^{+0.02}$ & $0.61400_{-0.02400}^{+0.02400}$ & $0.251_{-0.061}^{+0.061}$ & -- & $0.97_{-0.01}^{+0.01}$ & $1.119_{-0.033}^{+0.033}$ & $0.04294_{-0.00035}^{+0.00035}$ & $3.3777280_{-0.00001300}^{+0.00001300}$ & $9.00_{-1.70}^{+1.70}$ & e<0.069(<95 \%) & \cite{bakos2021hat} \\
HATS-5 b & $0.94_{-0.03}^{+0.03}$ & $0.23700_{-0.01200}^{+0.01200}$ & $0.190_{-0.080}^{+0.080}$ & -- & $0.87_{-0.02}^{+0.02}$ & $0.912_{-0.025}^{+0.025}$ & $0.05420_{-0.00060}^{+0.00060}$ & $4.7633870_{-0.00001000}^{+0.00001000}$ & $3.60_{-1.90}^{+2.60}$ & $0.0190_{-0.0190}^{+0.0190}$ & \cite{zhou2014hats} \\
HATS-8 b & $1.06_{-0.04}^{+0.04}$ & $0.13800_{-0.01900}^{+0.01900}$ & $0.210_{-0.080}^{+0.080}$ & -- & $1.09_{-0.06}^{+0.15}$ & $0.873_{-0.075}^{+0.123}$ & $0.04667_{-0.00055}^{+0.00055}$ & $3.5838930_{-0.00001000}^{+0.00001000}$ & $5.10_{-1.70}^{+1.70}$ & e<0.376(<95 \%) & \cite{bayliss2015hats} \\
HATS-11 b & $1.00_{-0.06}^{+0.06}$ & $0.85000_{-0.12000}^{+0.12000}$ & $-0.390_{-0.060}^{+0.060}$ & $1.1800$ & $1.44_{-0.06}^{+0.06}$ & $1.510_{-0.078}^{+0.078}$ & $0.04614_{-0.00093}^{+0.00093}$ & $3.6191613_{-0.00000990}^{+0.00000990}$ & $7.70_{-1.60}^{+2.20}$ & e<0.34(<95 \%) & \cite{rabus2016hats} \\
HATS-13 b & $0.96_{-0.03}^{+0.03}$ & $0.54300_{-0.07200}^{+0.07200}$ & $0.050_{-0.060}^{+0.060}$ & -- & $0.89_{-0.02}^{+0.02}$ & $1.212_{-0.035}^{+0.035}$ & $0.04057_{-0.00041}^{+0.00041}$ & $3.0440499_{-0.00000270}^{+0.00000270}$ & $2.50_{-1.70}^{+1.70}$ & e<0.181(<95 \%) & \cite{mancini2015hats} \\
HATS-14 b & $0.97_{-0.02}^{+0.02}$ & $1.07100_{-0.07000}^{+0.07000}$ & $0.330_{-0.060}^{+0.060}$ & -- & $0.93_{-0.01}^{+0.02}$ & $1.039_{-0.022}^{+0.032}$ & $0.03815_{-0.00032}^{+0.00032}$ & $2.7667641_{-0.00000270}^{+0.00000270}$ & $4.90_{-1.70}^{+1.70}$ & e<0.142(<95 \%) & \cite{mancini2015hats} \\
HATS-18 b & $1.04_{-0.05}^{+0.05}$ & $1.98000_{-0.07700}^{+0.07700}$ & $0.280_{-0.080}^{+0.080}$ & -- & $1.02_{-0.03}^{+0.06}$ & $1.337_{-0.049}^{+0.102}$ & $0.01761_{-0.00027}^{+0.00027}$ & $0.8378434_{-0.00000047}^{+0.00000047}$ & $4.20_{-2.20}^{+2.20}$ & e<0.166(<95 \%) & \cite{penev2016hats} \\
HATS-23 b & $1.12_{-0.05}^{+0.05}$ & $1.47000_{-0.07200}^{+0.07200}$ & $0.280_{-0.070}^{+0.070}$ & -- & $1.20_{-0.08}^{+0.06}$ & $1.860_{-0.400}^{+0.300}$ & $0.03397_{-0.00047}^{+0.00047}$ & $2.1605156_{-0.00000450}^{+0.00000450}$ & $4.20_{-1.50}^{+1.50}$ & e<0.114(<95 \%) & \cite{bento2017hats} \\
HATS-25 b & $0.99_{-0.04}^{+0.04}$ & $0.61300_{-0.04200}^{+0.04200}$ & $0.020_{-0.050}^{+0.050}$ & -- & $1.11_{-0.07}^{+0.07}$ & $1.260_{-0.100}^{+0.100}$ & $0.05163_{-0.00060}^{+0.00060}$ & $4.2986432_{-0.00000450}^{+0.00000450}$ & $7.50_{-1.90}^{+1.90}$ & e<0.176(<95 \%) & \cite{espinoza2016hats} \\
HATS-28 b & $0.93_{-0.04}^{+0.04}$ & $0.67200_{-0.08700}^{+0.08700}$ & $0.010_{-0.060}^{+0.060}$ & -- & $0.92_{-0.04}^{+0.04}$ & $1.194_{-0.070}^{+0.070}$ & $0.04131_{-0.00053}^{+0.00053}$ & $3.1810781_{-0.00000390}^{+0.00000390}$ & $6.20_{-2.80}^{+2.80}$ & e<0.202(<95 \%) & \cite{espinoza2016hats} \\
HATS-29 b & $1.03_{-0.05}^{+0.05}$ & $0.65300_{-0.06300}^{+0.06300}$ & $0.160_{-0.080}^{+0.080}$ & -- & $1.07_{-0.04}^{+0.04}$ & $1.251_{-0.061}^{+0.061}$ & $0.05475_{-0.00088}^{+0.00088}$ & $4.6058749_{-0.00000630}^{+0.00000630}$ & $5.50_{-1.70}^{+2.60}$ & e<0.158(<95 \%) & \cite{espinoza2016hats} \\
HATS-32 b & $1.10_{-0.04}^{+0.04}$ & $0.92000_{-0.10000}^{+0.10000}$ & $0.390_{-0.050}^{+0.050}$ & -- & $1.10_{-0.06}^{+0.10}$ & $1.249_{-0.096}^{+0.144}$ & $0.04024_{-0.00053}^{+0.00053}$ & $2.8126548_{-0.00000550}^{+0.00000550}$ & $3.50_{-1.80}^{+1.80}$ & e<0.471(<95 \%) & \cite{de2016hats} \\
HATS-33 b & $1.06_{-0.03}^{+0.03}$ & $1.19200_{-0.05300}^{+0.05300}$ & $0.290_{-0.050}^{+0.050}$ & -- & $1.02_{-0.04}^{+0.05}$ & $1.230_{-0.081}^{+0.112}$ & $0.03727_{-0.00037}^{+0.00037}$ & $2.5495551_{-0.00000610}^{+0.00000610}$ & $3.00_{-1.70}^{+1.70}$ & e<0.08(<95 \%) & \cite{de2016hats} \\
HATS-34 b & $0.95_{-0.03}^{+0.03}$ & $0.94100_{-0.07200}^{+0.07200}$ & $0.250_{-0.070}^{+0.070}$ & -- & $0.98_{-0.05}^{+0.05}$ & $1.430_{-0.190}^{+0.190}$ & $0.03166_{-0.00034}^{+0.00034}$ & $2.1061607_{-0.00000470}^{+0.00000470}$ & $7.70_{-2.70}^{+2.70}$ & e<0.108(<95 \%) & \cite{de2016hats} \\
HATS-43 b & $0.84_{-0.02}^{+0.02}$ & $0.26100_{-0.05400}^{+0.05400}$ & $0.050_{-0.041}^{+0.041}$ & -- & $0.81_{-0.03}^{+0.03}$ & $1.180_{-0.050}^{+0.050}$ & $0.04944_{-0.00046}^{+0.00046}$ & $4.3888497_{-0.00000590}^{+0.00000590}$ & $8.60_{-4.80}^{+3.00}$ & $0.1730_{-0.0890}^{+0.0890}$ & \cite{brahm2018hats} \\
HATS-44 b & $0.86_{-0.02}^{+0.02}$ & $0.56000_{-0.11000}^{+0.11000}$ & $0.320_{-0.071}^{+0.071}$ & -- & $0.85_{-0.04}^{+0.04}$ & $1.067_{-0.071}^{+0.125}$ & $0.03649_{-0.00030}^{+0.00030}$ & $2.7439004_{-0.00000320}^{+0.00000320}$ & $9.70_{-4.00}^{+2.40}$ & e<0.279(<95 \%) & \cite{brahm2018hats} \\
HATS-47 b & $0.67_{-0.01}^{+0.02}$ & $0.36900_{-0.02100}^{+0.03100}$ & $-0.113_{-0.035}^{+0.035}$ & -- & $0.66_{-0.01}^{+0.01}$ & $1.117_{-0.014}^{+0.014}$ & $0.04269_{-0.00025}^{+0.00033}$ & $3.9228038_{-0.00000220}^{+0.00000220}$ & $8.10_{-4.30}^{+2.90}$ & e<0.088(<95 \%) & \cite{hartman2020hats} \\
HATS-54 b & $1.10_{-0.02}^{+0.02}$ & $0.76000_{-0.10000}^{+0.10000}$ & $0.396_{-0.031}^{+0.031}$ & -- & $1.32_{-0.04}^{+0.04}$ & $1.067_{-0.052}^{+0.052}$ & $0.03763_{-0.00024}^{+0.00024}$ & $2.5441828_{-0.00000430}^{+0.00000430}$ & $6.60_{-0.76}^{+0.76}$ & e<0.126(<95 \%) & \cite{espinoza2019hats} \\
HATS-57 b & $1.03_{-0.03}^{+0.02}$ & $3.14700_{-0.07300}^{+0.07300}$ & $0.268_{-0.043}^{+0.043}$ & -- & $0.96_{-0.01}^{+0.01}$ & $1.139_{-0.028}^{+0.028}$ & $0.03493_{-0.00030}^{+0.00021}$ & $2.3506210_{-0.00000130}^{+0.00000130}$ & $2.50_{-1.10}^{+1.50}$ & e<0.028(<95 \%) & \cite{espinoza2019hats} \\
HATS-76 b & $0.66_{-0.02}^{+0.02}$ & $2.62900_{-0.08900}^{+0.08900}$ & $0.322_{-0.049}^{+0.065}$ & -- & $0.63_{-0.01}^{+0.01}$ & $1.079_{-0.031}^{+0.031}$ & $0.02658_{-0.00029}^{+0.00021}$ & $1.9416423_{-0.00000140}^{+0.00000140}$ & $4.60_{-4.00}^{+8.70}$ & e<0.062(<95 \%) & \cite{jordan2022hats} \\
K2-30 b & $0.92_{-0.01}^{+0.01}$ & $0.58900_{-0.02200}^{+0.02300}$ & $0.060_{-0.040}^{+0.040}$ & -- & $0.84_{-0.01}^{+0.02}$ & $1.069_{-0.019}^{+0.023}$ & $0.04190_{-0.00120}^{+0.00160}$ & $4.0984900_{-0.00002000}^{+0.00002000}$ & $2.20_{-0.80}^{+1.80}$ & e<0.08(<96 \%) & \cite{brahm2016independent} \\
K2-132 b & $1.19_{-0.04}^{+0.04}$ & $0.49500_{-0.00630}^{+0.00680}$ & $-0.110_{-0.050}^{+0.050}$ & -- & $4.11_{-0.05}^{+0.05}$ & $1.089_{-0.006}^{+0.006}$ & $0.09160_{-0.00060}^{+0.00060}$ & $9.1708000_{-0.00250000}^{+0.00250000}$ & $5.50_{-0.40}^{+0.40}$ & $0.2900_{-0.0490}^{+0.0490}$ & \cite{jones2018hot} \\
K2-140 b & $1.00_{-0.02}^{+0.02}$ & $1.01900_{-0.07000}^{+0.07000}$ & $0.120_{-0.045}^{+0.045}$ & -- & $0.99_{-0.01}^{+0.01}$ & $1.095_{-0.018}^{+0.018}$ & $0.05910_{-0.00340}^{+0.00340}$ & $6.5693000_{-0.00002000}^{+0.00001700}$ & $4.22_{-0.95}^{+0.95}$ & $0.1200_{-0.0460}^{+0.0560}$ & \cite{giles2018k2} \\
Kepler-12 b & $1.17_{-0.05}^{+0.05}$ & $0.43100_{-0.04000}^{+0.04100}$ & $0.070_{-0.040}^{+0.040}$ & $1.3765_{-0.0020}^{+0.0020}$ & $1.48_{-0.03}^{+0.03}$ & $1.695_{-0.032}^{+0.028}$ & $0.05560_{-0.00070}^{+0.00070}$ & $4.4379637_{-0.00000020}^{+0.00000020}$ & $4.00_{-0.40}^{+0.30}$ & e<0.01(<84.1 \%) & \cite{fortney2011discovery} \\
Kepler-44 b & $1.19_{-0.10}^{+0.10}$ & $1.02000_{-0.07000}^{+0.07000}$ & $0.260_{-0.100}^{+0.100}$ & -- & $1.52_{-0.09}^{+0.09}$ & $1.240_{-0.070}^{+0.070}$ & $0.04550_{-0.00130}^{+0.00130}$ & $3.2467400_{-0.00001800}^{+0.00001800}$ & $6.95_{-1.70}^{+1.10}$ & e<0.021(<84.1 \%) & \cite{bonomo2012sophie} \\
Kepler-412 b & $1.17_{-0.09}^{+0.09}$ & $0.93900_{-0.08500}^{+0.08500}$ & $0.270_{-0.120}^{+0.120}$ & -- & $1.29_{-0.04}^{+0.04}$ & $1.325_{-0.043}^{+0.043}$ & $0.02959_{-0.00078}^{+0.00078}$ & $1.7208612_{-0.00000005}^{+0.00000005}$ & $5.10_{-1.70}^{+1.70}$ & $0.0038_{-0.0032}^{+0.0087}$ & \cite{deleuil2014sophie} \\
Kepler-425 b & $0.93_{-0.05}^{+0.05}$ & $0.25000_{-0.08000}^{+0.08000}$ & $0.240_{-0.110}^{+0.110}$ & -- & $0.86_{-0.02}^{+0.02}$ & $0.978_{-0.022}^{+0.022}$ & $0.04640_{-0.00080}^{+0.00080}$ & $3.7970182_{-0.00000019}^{+0.00000019}$ & $5.00_{-4.00}^{+4.00}$ & e<0.33(<99 \%) & \cite{hebrard2014characterization} \\
Kepler-426 b & $0.91_{-0.06}^{+0.06}$ & $0.34000_{-0.08000}^{+0.08000}$ & $-0.210_{-0.080}^{+0.080}$ & -- & $0.92_{-0.02}^{+0.02}$ & $1.090_{-0.030}^{+0.030}$ & $0.04140_{-0.00100}^{+0.00100}$ & $3.2175188_{-0.00000019}^{+0.00000019}$ & $6.00_{-4.00}^{+4.00}$ & e<0.18(<99 \%) & \cite{hebrard2014characterization} \\
Kepler-428 b & $0.87_{-0.05}^{+0.05}$ & $1.27000_{-0.19000}^{+0.19000}$ & $0.090_{-0.170}^{+0.170}$ & -- & $0.80_{-0.02}^{+0.02}$ & $1.080_{-0.030}^{+0.030}$ & $0.04330_{-0.00090}^{+0.00090}$ & $3.5256325_{-0.00000015}^{+0.00000015}$ & $5.00_{-4.00}^{+4.00}$ & e<0.22(<99 \%) & \cite{hebrard2014characterization} \\
TOI-172 b & $1.13_{-0.06}^{+0.07}$ & $5.42000_{-0.20000}^{+0.22000}$ & $0.148_{-0.080}^{+0.079}$ & $0.3120_{-0.0100}^{+0.0100}$ & $1.78_{-0.04}^{+0.05}$ & $0.965_{-0.029}^{+0.032}$ & $0.09140_{-0.00170}^{+0.00170}$ & $9.4772500_{-0.00079000}^{+0.00064000}$ & $7.40_{-1.50}^{+1.60}$ & $0.3806_{-0.0090}^{+0.0093}$ & \cite{rodriguez2019eccentric} \\
TOI-559 b & $1.03_{-0.06}^{+0.06}$ & $6.01000_{-0.23000}^{+0.24000}$ & $-0.069_{-0.079}^{+0.065}$ & $0.8276_{-0.0090}^{+0.0100}$ & $1.23_{-0.03}^{+0.03}$ & $1.091_{-0.025}^{+0.028}$ & $0.07230_{-0.00130}^{+0.00130}$ & $6.9839095_{-0.00000510}^{+0.00000510}$ & $6.80_{-2.00}^{+2.50}$ & $0.1510_{-0.0110}^{+0.0120}$ & \cite{ikwut2021two} \\
TOI-564 b & $1.00_{-0.06}^{+0.07}$ & $1.46300_{-0.09600}^{+0.10000}$ & $0.143_{-0.078}^{+0.076}$ & $0.9200_{-0.4500}^{+1.7000}$ & $1.09_{-0.01}^{+0.01}$ & $1.020_{-0.290}^{+0.710}$ & $0.02734_{-0.00053}^{+0.00061}$ & $1.6511440_{-0.00001800}^{+0.00001800}$ & $7.30_{-3.50}^{+3.60}$ & $0.0720_{-0.0500}^{+0.0830}$ & \cite{davis2020toi} \\
TOI-905 b & $0.97_{-0.07}^{+0.06}$ & $0.66700_{-0.04100}^{+0.04200}$ & $0.140_{-0.180}^{+0.220}$ & $1.7180_{-0.0300}^{+0.0320}$ & $0.92_{-0.04}^{+0.04}$ & $1.171_{-0.051}^{+0.053}$ & $0.04666_{-0.00110}^{+0.00096}$ & $3.7394940_{-0.00003800}^{+0.00003800}$ & $3.40_{-2.30}^{+3.80}$ & $0.0240_{-0.0170}^{+0.0250}$ & \cite{davis2020toi} \\
TOI-1268 b & $0.96_{-0.04}^{+0.04}$ & $0.30331_{-0.02611}^{+0.02580}$ & $0.360_{-0.060}^{+0.060}$ & $0.8186_{-0.0192}^{+0.0205}$ & $0.92_{-0.06}^{+0.06}$ & $0.812_{-0.054}^{+0.054}$ & $0.07110_{-0.00580}^{+0.00630}$ & $8.1577094_{-0.00000450}^{+0.00000450}$ & $0.24_{-0.14}^{+0.14}$ & $0.1050_{-0.0390}^{+0.0400}$ & \cite{vsubjak2022toi} \\
TOI-1296 b & $1.17_{-0.14}^{+0.14}$ & $0.29800_{-0.03900}^{+0.03900}$ & $0.440_{-0.040}^{+0.040}$ & -- & $1.66_{-0.04}^{+0.04}$ & $1.231_{-0.031}^{+0.031}$ & $0.04970_{-0.00280}^{+0.00230}$ & $3.9443715_{-0.00000580}^{+0.00000580}$ & $6.90_{-0.70}^{+0.70}$ & $0.0550_{-0.0380}^{+0.0610}$ & \cite{moutou2021toi} \\
TOI-2158 b & $1.12_{-0.12}^{+0.12}$ & $0.82000_{-0.08000}^{+0.08000}$ & $0.470_{-0.080}^{+0.080}$ & -- & $1.41_{-0.03}^{+0.03}$ & $0.960_{-0.012}^{+0.012}$ & $0.07500_{-0.00400}^{+0.00400}$ & $8.6007700_{-0.00003000}^{+0.00003000}$ & $8.00_{-1.00}^{+1.00}$ & $0.0310_{-0.0310}^{+0.0130}$ & \cite{knudstrup2022confirmation} \\
TOI-2669 b & $1.19_{-0.16}^{+0.16}$ & $0.61000_{-0.19000}^{+0.19000}$ & $0.100_{-0.060}^{+0.060}$ & -- & $4.10_{-0.04}^{+0.04}$ & $1.760_{-0.160}^{+0.160}$ & $0.00000_{0.00000}^{+0.00000}$ & $6.2034000_{-0.00010000}^{+0.00010000}$ & $5.90_{-3.00}^{+3.00}$ & $0.0900_{-0.0500}^{+0.0500}$ & \cite{grunblatt2022tess} \\
TOI-3629 b & $0.63_{-0.02}^{+0.02}$ & $0.26000_{-0.02000}^{+0.02000}$ & $0.400_{-0.100}^{+0.100}$ & -- & $0.60_{-0.01}^{+0.02}$ & $0.740_{-0.020}^{+0.020}$ & $0.04300_{-0.00200}^{+0.00200}$ & $3.9365510_{-0.00000600}^{+0.00000500}$ & $7.00_{-2.00}^{+2.00}$ & $0.0500_{-0.0400}^{+0.0500}$ & \cite{canas2022toi} \\
TOI-3757 b & $0.64_{-0.02}^{+0.02}$ & $0.26838_{-0.02737}^{+0.02769}$ & $0.000_{-0.200}^{+0.200}$ & -- & $0.62_{-0.01}^{+0.01}$ & $1.071_{-0.045}^{+0.036}$ & $0.03845_{-0.00043}^{+0.00043}$ & $3.4387530_{-0.00000400}^{+0.00000400}$ & $7.10_{-4.50}^{+4.50}$ & $0.1400_{-0.0600}^{+0.0600}$ & \cite{kanodia2022toi} \\
TrES-3 b & $0.93_{-0.05}^{+0.03}$ & $1.80100_{-0.08300}^{+0.07800}$ & $-0.190_{-0.080}^{+0.080}$ & -- & $0.83_{-0.02}^{+0.01}$ & $1.336_{-0.037}^{+0.031}$ & $0.02284_{-0.00037}^{+0.00024}$ & $1.3061865_{-0.00000007}^{+0.00000007}$ & $0.90_{-0.80}^{+2.80}$ & e<0.043(<84.1 \%) & \cite{bonomo2017gaps} \\
WASP-4 b & $0.89_{-0.01}^{+0.01}$ & $1.17000_{-0.01500}^{+0.01500}$ & $-0.050_{-0.040}^{+0.040}$ & -- & $0.92_{-0.06}^{+0.06}$ & $1.330_{-0.160}^{+0.160}$ & $0.02287_{-0.00009}^{+0.00009}$ & $1.3382316_{-0.00000007}^{+0.00000007}$ & $7.00_{-2.90}^{+2.90}$ & e<0.0033(<84.1 \%) & \cite{bonomo2017gaps} \\
WASP-5 b & $0.96_{-0.09}^{+0.13}$ & $1.58000_{-0.10000}^{+0.13000}$ & $0.090_{-0.090}^{+0.090}$ & $1.1800_{-0.0290}^{+0.0220}$ & $1.03_{-0.07}^{+0.06}$ & $1.087_{-0.071}^{+0.068}$ & $0.02670_{-0.00080}^{+0.00120}$ & $1.6284279_{-0.00000490}^{+0.00000220}$ & $5.40_{-4.30}^{+4.40}$ & $0.0380_{-0.0180}^{+0.0260}$ & \cite{gillon2009improved} \\
WASP-6 b & $0.84_{-0.06}^{+0.06}$ & $0.48500_{-0.02700}^{+0.02700}$ & $-0.150_{-0.090}^{+0.090}$ & -- & $0.86_{-0.02}^{+0.02}$ & $1.230_{-0.035}^{+0.035}$ & $0.04140_{-0.00100}^{+0.00100}$ & $3.3610021_{-0.00000031}^{+0.00000031}$ & $9.00_{-12.70}^{+8.00}$ & $0.0540_{-0.0150}^{+0.0180}$ & \cite{tregloan2015transits} \\
WASP-19 b & $0.96_{-0.10}^{+0.09}$ & $1.15400_{-0.08000}^{+0.07800}$ & $0.040_{-0.300}^{+0.250}$ & $2.0770_{-0.0140}^{+0.0140}$ & $1.01_{-0.03}^{+0.03}$ & $1.415_{-0.048}^{+0.044}$ & $0.01652_{-0.00056}^{+0.00050}$ & $0.7888385_{-0.00000082}^{+0.00000075}$ & $6.40_{-3.50}^{+4.10}$ & $0.0126_{-0.0089}^{+0.0140}$ & \cite{bonomo2017gaps} \\
WASP-22 b & $1.10_{-0.30}^{+0.30}$ & $0.56000_{-0.02000}^{+0.02000}$ & $-0.050_{-0.080}^{+0.080}$ & $1.0400_{-0.0400}^{+0.0400}$ & $1.13_{-0.03}^{+0.03}$ & $1.120_{-0.040}^{+0.040}$ & $0.04680_{-0.00040}^{+0.00040}$ & $3.5326900_{-0.00004000}^{+0.00004000}$ & $3.00_{-1.00}^{+1.00}$ & $0.0230_{-0.0120}^{+0.0120}$ & \cite{maxted2010wasp} \\
WASP-23 b & $0.78_{-0.12}^{+0.13}$ & $0.87900_{-0.10000}^{+0.09500}$ & $-0.050_{-0.130}^{+0.130}$ & -- & $0.77_{-0.05}^{+0.03}$ & $0.962_{-0.056}^{+0.047}$ & $0.03700_{-0.00220}^{+0.00190}$ & $2.9444300_{-0.00000110}^{+0.00000110}$ & $6.20_{-2.50}^{+5.60}$ & e<0.065(<84.1 \%) & \cite{bonomo2017gaps} \\
WASP-32 b & $1.07_{-0.05}^{+0.05}$ & $3.49000_{-0.11000}^{+0.12000}$ & $-0.130_{-0.100}^{+0.100}$ & -- & $1.09_{-0.03}^{+0.03}$ & $1.100_{-0.040}^{+0.040}$ & $0.03904_{-0.00061}^{+0.00062}$ & $2.7186591_{-0.00000240}^{+0.00000240}$ & $2.22_{-0.73}^{+0.62}$ & e<0.004(<84.1 \%) & \cite{bonomo2017gaps} \\
WASP-43 b & $0.72_{-0.03}^{+0.03}$ & $2.05000_{-0.05200}^{+0.05000}$ & $-0.050_{-0.170}^{+0.170}$ & -- & $0.67_{-0.01}^{+0.01}$ & $1.036_{-0.019}^{+0.019}$ & $0.01528_{-0.00018}^{+0.00017}$ & $0.8134744_{-0.00000013}^{+0.00000013}$ & $7.00_{-7.00}^{+7.00}$ & e<0.0062(<84.1 \%) & \cite{bonomo2017gaps} \\
WASP-50 b & $0.89_{-0.07}^{+0.08}$ & $1.46000_{-0.08700}^{+0.08700}$ & $-0.120_{-0.080}^{+0.080}$ & -- & $0.84_{-0.03}^{+0.03}$ & $1.153_{-0.048}^{+0.048}$ & $0.02945_{-0.00087}^{+0.00084}$ & $1.9550938_{-0.00000130}^{+0.00000130}$ & $7.00_{-3.50}^{+3.50}$ & e<0.022(<84.1 \%) & \cite{bonomo2017gaps} \\
WASP-59 b & $0.72_{-0.04}^{+0.04}$ & $0.86300_{-0.04500}^{+0.04500}$ & $-0.150_{-0.110}^{+0.110}$ & $1.6900_{-0.0800}^{+0.0800}$ & $0.61_{-0.04}^{+0.04}$ & $0.775_{-0.068}^{+0.068}$ & $0.06970_{-0.00110}^{+0.00110}$ & $7.9195850_{-0.00001000}^{+0.00001000}$ & $0.50_{-0.40}^{+0.70}$ & $0.1000_{-0.0420}^{+0.0420}$ & \cite{hebrard2013wasp} \\
WASP-89 b & $0.92_{-0.08}^{+0.08}$ & $5.90000_{-0.40000}^{+0.40000}$ & $0.150_{-0.140}^{+0.140}$ & $1.4900_{-0.0200}^{+0.0200}$ & $0.88_{-0.03}^{+0.03}$ & $1.040_{-0.040}^{+0.040}$ & $0.04270_{-0.00120}^{+0.00120}$ & $3.3564227_{-0.00000250}^{+0.00000250}$ & $1.30_{-0.80}^{+1.50}$ & $0.1930_{-0.0090}^{+0.0090}$ & \cite{hellier2015three} \\
WASP-119 b & $1.02_{-0.06}^{+0.06}$ & $1.23000_{-0.08000}^{+0.08000}$ & $0.140_{-0.100}^{+0.100}$ & -- & $1.20_{-0.10}^{+0.10}$ & $1.400_{-0.200}^{+0.200}$ & $0.03630_{-0.00070}^{+0.00070}$ & $2.4997900_{-0.00001000}^{+0.00001000}$ & $8.00_{-2.50}^{+2.50}$ & e<0.058(<95 \%) & \cite{maxted2016five} \\
WASP-124 b & $1.07_{-0.05}^{+0.05}$ & $0.60000_{-0.07000}^{+0.07000}$ & $-0.020_{-0.110}^{+0.110}$ & -- & $1.02_{-0.02}^{+0.02}$ & $1.240_{-0.030}^{+0.030}$ & $0.04490_{-0.00070}^{+0.00070}$ & $3.3726500_{-0.00000100}^{+0.00000100}$ & $2.10_{-1.40}^{+1.40}$ & e<0.017(<95 \%) & \cite{maxted2016five} \\
WASP-133 b & $1.16_{-0.08}^{+0.08}$ & $1.16000_{-0.09000}^{+0.09000}$ & $0.290_{-0.120}^{+0.120}$ & -- & $1.44_{-0.05}^{+0.05}$ & $1.210_{-0.050}^{+0.050}$ & $0.03450_{-0.00070}^{+0.00070}$ & $2.1764230_{-0.00000100}^{+0.00000100}$ & $6.80_{-1.80}^{+1.80}$ & e<0.17(<95 \%) & \cite{maxted2016five} \\
WASP-151 b & $1.08_{-0.08}^{+0.08}$ & $0.31000_{-0.03000}^{+0.04000}$ & $0.100_{-0.100}^{+0.100}$ & -- & $1.14_{-0.03}^{+0.03}$ & $1.130_{-0.030}^{+0.030}$ & $0.05500_{-0.00100}^{+0.00100}$ & $4.5334710_{-0.00000400}^{+0.00000400}$ & $5.10_{-1.30}^{+1.30}$ & e<0.003(<84.1 \%) & \cite{demangeon2018discovery} \\
WASP-185 b & $1.12_{-0.06}^{+0.06}$ & $0.98000_{-0.06000}^{+0.06000}$ & $-0.020_{-0.060}^{+0.060}$ & $0.7300_{-0.0500}^{+0.0500}$ & $1.50_{-0.08}^{+0.08}$ & $1.250_{-0.080}^{+0.080}$ & $0.09040_{-0.00170}^{+0.00170}$ & $9.3875500_{-0.00002000}^{+0.00002000}$ & $6.60_{-1.60}^{+1.60}$ & $0.2400_{-0.0400}^{+0.0400}$ & \cite{hellier2019wasp} \\
\end{tabular}
}
\end{table*}

We select exoplanet systems for analysis from the NASA Exoplanet Archive
(\url{https://exoplanetarchive.ipac.caltech.edu/}) using the following filters:

\begin{enumerate}
    \item Single planet, single star transiting systems
    \item Orbital period less than 10 days
    \item Planet radius between 0.6 and 1.2 Jupiter radii
    \item Planet mass less than 13 Jupiter masses
    \item Stellar mass between 0.4 and 1.2 Solar masses
    \item Nominal eccentricity (if given) or eccentricity upper limit if only
        that is available less than 0.5
    \item Stellar metallicity in the range $-1.014 < \feh < 0.537$
\end{enumerate}

The first condition aims to exclude systems for which the size of the planet is
unknown or highly uncertain, or where three-body interactions may dominate over
the tidal effects on the orbit. The second condition selects systems where tides
are strong enough to affect the orbit. The third and fourth conditions ensure
our sample includes only gas giant planets so that we can reasonably assume
the same tidal dissipation will apply to all planets in our sample.
%
%
    This assumption may break down, for example, if the cores of planets vary
    widely, or if variable instellation changes the surface layers sufficiently
    to affect the dissipation, or if variations in metallicity or rotation
    affect the internal structure and hence the dissipation. Ultimately, we find
    that a common $Q_{pl}'$ prescription is capable of explaining the observed
    circularization of all planets in our sample, so we do not see indications
    of such effects at least in this sample of planets, though of course we
    cannot rule them out.
%
%
The fifth condition requires stars with significant surface convective zones,
ensuring the effects of stellar tides will be similar. The sixth condition is to
ensure the assumptions behind our method of constraining tidal dissipation are
valid. In particular, when we estimate tidal dissipation we assume that initial
eccentricity, before tidal evolution began, must have been 0.8 or below. This is
guided by a number of theoretical and observational lines of evidence that
extremely high orbital eccentricities may be damped very quickly by some process
in the planet that becomes ineffective once the eccentricity is no longer close
to unity \citep[e.g.][]{Wu_18, Moe_Krater_18, Mardling_95}. However, since that
is not a fully understood process, we wish to exclude systems that may have
avoided this fast pre-circularization. These would be the hot Jupiters with the
highest observed present-day eccentricities. The final condition is imposed by
our tool for calculating orbital evolutions, called POET \citep{penev2014poet},
which accounts for stellar evolution by interpolating within a grid of stellar
evolution tracks which are limited to the given metallicity range.

This selection resulted in 78 hot Jupiter systems selected for analysis. Table
\ref{tab:exotable} lists the relevant properties of the exoplanet systems
adopted for this analysis as well the source references for these parameters.
The second to last column of the table shows the orbital eccentricities of the
systems. If the maximum limit of the eccentricity ($e_{max}$) is reported in the
reference paper with a $c\%$ confidence, eccentricity is written in the column in
the format of `$e<e_{max}(<c\%)$'. If the median value of the eccentricity ($e$)
is reported in the paper with the upper and lower uncertainties ($+\sigma_{u}$
and $-\sigma_{l}$, respectively) up to 1-$\sigma$ interval, then the
eccentricity is written in the format of ${e}^{+\sigma_{u}}_{-\sigma_{l}}$. In
either case, when defining our likelihood function for the analysis we use the
confidence level specified in the relevant paper (see section
\ref{sec:bayesianAnalysis}).


One of the 78 systems, CoRoT-23 b, is the subject of some disagreement in the
literature. Its eccentricity is reported to be $0.16^{+0.02}_{-0.02}$ in
\cite{rouan2012transiting}. However, later studies by \cite{bonomo2017gaps}
showed that the nonzero eccentricity of this system's orbit is dubious. We have
not included this system while calculating the combined constraints for all
systems, but we did extract individual constraints on $Q_{pl}'$ for that system
following the same procedure as the other planets in our sample; these are also
reported in Section \ref{sec:individual_constraints}.

Our sample also includes two systems with red giant primary stars: K2-132 and TOI-2669. For those systems, the dissipation in the star likely dominates the present-day orbital evolution and we suspect the way stellar evolution is handled in our modeling may not be quite adequate to handle the rapidly evolving stellar radii. Just like for CoRoT-23, we report individual analyses of the tides in these systems but do not include them when constructing the combined constraints.

\section{Methodology}

\subsection{Tidal Evolution Model}
\label{sec:evolution}

To calculate the tidal orbital evolution, we follow the same formalism as
\citet{penev2014poet, patel2022constraining, penev2022comprehensive}. This
formalism is an extension of \citet{lai2012tidal} for eccentric orbits. The
evolution calculations are performed by an updated version of the publicly
available tidal evolution module \textbf{P}lanerary \textbf{O}rbital
\textbf{E}volution due to \textbf{T}ides (POET hereafter;
\citet{penev2014poet}).  Below, we give a brief outline of the POET formalism
for completeness.

When two bodies of mass $M$ and $M'$ revolve around their barycenter, one is
subject to the tidal force of the other.  The tidal potential associated with
the gravitational field created by $M'$ at the position vector
$\textbf{r}$ with respect to the center of mass of $M$ is given by:
\begin{equation}
    U(\mathbf{r}, t) = \frac{GM'}{|\mathbf{r}_{M'}|}
    \left(
        \frac{\mathbf{r}\cdot\mathbf{r}_{M'}}{\left|\mathbf{r}_{M'}\right|^2}
        -
        \frac{\left|\mathbf{r_{M'}}\right|}{\left|\mathbf{r}-\mathbf{r}_{M'}\right|}
    \right)
    \label{tidalpotentialequation}
\end{equation}
Here, $\textbf{r}_{M'}$ is the position vector of the center of the mass $M'$
from the center of the mass $M$.

If $a$ is the semi-major axis of the orbit, 
$\textbf{r}$, Equation \ref{tidalpotentialequation} can be expanded up to the
lowest order of $\frac{\left|\mathbf{r}\right|}{a}$ in terms of spherical
harmonics as follows:
\begin{equation}
    \begin{split} U(\mathbf{r}, t) = & -\frac{GM'}{a}
        \left(\frac{\rho}{a}\right)^2
        \left(\frac{a}{\left|\mathbf{r_{M'}}\right|}\right)^3\\
        &\sum_{m,m'}
        W_{2,m'} \mathcal{D}_{m,m'}(\Theta) Y_{2,m}(\theta,\phi)
        \exp\left(-im'\Delta \Phi \right)
    \end{split}
\end{equation}
where $\Theta$ is the angle between the orbital angular momentum
($\textbf{\textit{L}}$) and the spin angular momentum of $M$
($\textbf{\textit{S}}$), i.e. the obliquity of the mass $M$, $\Delta\Phi$ is
the true anomaly of $M'$ in its orbit around $M$, and ($\rho$, $\theta$, $\phi$)
are the spherical polar coordinates of $\textbf{r}$ with respect to the axes
system having an origin at the center of the mass $M$, $z$-axis in the direction
of $\textbf{\textit{S}}$ and $y$-axis in the direction of $\textbf{\textit{S}}
\times \textbf{\textit{L}}$. Finally, $\mathcal{D}_{m,m'}(\Theta)$ are the
elements of Wigner $\mathcal{D}$ matrix of $l=2$, $W_{2,0} = -\sqrt{\pi/5}$ and
$W_{2,\pm 2} = \sqrt{3\pi/10}$.

Since each coefficient in the above expansion is time-dependent, we expand as a
Fourier series in time:
\begin{equation}
    \frac{a^3}{\left|\mathbf{r_{M'}}\right|^3} \exp\left(-im'\Delta \Phi \right)
    =
    \sum_{s} p_{m',s} e^{-is\Omega t}
    \label{eq:fourier}
\end{equation}

Here, $\Omega$ is the mean motion of $M'$ in its orbit around $M$, and
$p_{m',s}$ are expansion coefficients that depend solely on the orbital
eccentricity.
%
%
    We compute those numerically using the Fourier transform of Eq.
%
%
\ref{eq:fourier}.
%
%
    The summation is truncated to a maximum value of $s$ that is adjusted
    dynamically during the evolution to ensure the tidal potential is
    approximated to better than one part in $10^5$. Thus, when the eccentricity
    is high, the maximum $s$ included exceeds 100 but quickly drops to just
    several as eccentricity decreases.
%
%
We have pre-computed $p_{m',s}$ for a grid of eccentricities dense enough to
allow precise interpolation for any eccentricity.

Following \citet{lai2012tidal}, we postulate that the Lagrangian displacement
each tidal potential term will produce will be independent and lag slightly
behind the corresponding tidal potential term:
\begin{equation}
    U_{m,s} \equiv \frac{GM'}{a^3} \sum_m' W_{2,m'}D_{m,m'}(\Theta) p_{m',s}
\end{equation}
thus:
\begin{equation}
    \boldsymbol{\xi}_{m,s}(\mathbf{r},t)
    =
    \frac{U_{m,s}}{\omega_0^2}\bar{\boldsymbol{\xi}}_{m,s}(\mathbf{r})
    \exp(-i\tilde{\omega}_{m,s} t + i\Delta_{m,s})
    \label{eq:displacement}
\end{equation}
and consequently, the density perturbation will be:
\begin{equation}
    \delta\rho_{m,s}(\mathbf{r},t)
    =
    \frac{U_{m,s}}{\omega_0^2} \delta\bar{\rho}_{m,s}(\mathbf{r})
    \exp(-i\tilde{\omega}_{m,s} t + i\Delta_{m,s})
    \label{eq:density}
\end{equation}
with
\begin{equation}
    \delta\bar{\rho}_{m,s}=-\nabla\cdot(\rho\bar{\boldsymbol{\xi}}_{m,s})
\end{equation}
Here, $\omega_{0}\equiv\sqrt{GM/R^3}$ is the dynamic frequency of the object
experiencing the tides, $R$ is the radius of the body of mass $M$, $\Delta_{m,
s}$ is the phase lag between the tidal bulge and the tidal potential due to the
tidal dissipation, and $\tilde{\omega}_{m,s}$ is the tidal frequency which
includes the rotation of the star and is defined by:
\begin{equation}
    \tilde{\omega}_{m,s}=s\Omega-m\Omega_{spin}
\end{equation}

It is important to note that Eq.~\ref{eq:displacement} and \ref{eq:density} are
general enough to represent both equilibrium and dynamical tides as long as the
phase lag is allowed to be different for each tidal wave and evolves as the
tidally distorted object evolves. Each tidal model will provide a different
prescription for $\Delta_{m, s}$.

We make further assumptions that the tidal phase lag, $\Delta_{m,s}$, will only
depend on the tidal frequency and this dependence will be continuous.
where $\Omega_{spin}$ is the rotational angular speed of $M$ around the
$z$-axis.

From the above, we can calculate the tidal torque as:
\begin{equation}
    \textbf{\textit{T}}
    =
    -\int d^3 x \delta\rho(\mathbf{r}, t) \mathbf{r}\times \nabla
    U^*(\mathbf{r}, t)
    \label{eq:tidal_torque}
\end{equation}
and the energy transfer rate from the orbit to the body as:
\begin{equation}
    \dot{E}
    =
    -\int d^3 x \rho(\mathbf{r}) \frac{\partial \mathbf{\xi}(\mathbf{r},
    t)}{\partial t} \cdot\nabla U^*(\mathbf{r}, t)
    \label{eq:tidal_power}
\end{equation}
with
\begin{equation}
    \delta\rho(\mathbf{r}, t) = \sum_{m,s}\delta\rho_{m,s}(\mathbf{r}, t)
\end{equation}
\begin{equation}
    \mathbf{\xi}(\mathbf{r}, t) = \sum_{m,s}\mathbf{\xi}_{m,s}(\mathbf{r}, t)
\end{equation}
Here, * denotes complex conjugation.

Thanks to the orthonormality of the spherical harmonic and Fourier functions,
only terms where the indices of the potential match the indices of the
perturbations contribute.

With the further simplification that the obliquity is zero, i.e.  $\Theta = 0$,
the semi-major axis and the eccentricity of the orbit evolve with time according
to the following equations:
\begin{equation}
    \dot{a}=a\frac{-\dot{E}}{E}
\end{equation}

\begin{equation}
    \dot{e}=\frac{(M+M')}{G^2(MM')^3}\frac{(\dot{E}L+2E\dot{L})L}{e}
\end{equation}
where $\dot{E}$ and $\dot{L}$ include contributions from both tides on the star
and the planet.

Stars have an internal structure that also evolves with time. We restrict our
analysis to systems with parent star masses between 0.4 and 1.2 solar masses.
These stars have a radiative core and a convective envelope. We assume that
tides only couple to the convective region of the star, allowing the envelope to
directly exchange angular momentum with the orbit. The convective envelope then
couples to the radiative core as well as to a magnetized stellar wind.
%
%
    Note that several theoretical studies suggest that tidal dissipation in the
    radiative zones of exoplanet host stars may dominate over the dissipation in
    the convective zone.
%
%
\citep[c.f.][and others]{goodman1998dynamical, Ogilvie_Lin_07,
Barker_Ogilvie_10, Essick_Weinberg_16, Ivanov_et_al_13, barker2020tidal}
%
%
    However, that assumption is unlikely to have an important effect on our
    final results, since we already account for several orders of magnitude
    uncertainty in the efficiency of stellar tides (see Section
%
%
\ref{sec:q_prescription}).
%
%
    Compared to that, the difference between coupling to the core vs. the
    envelope is insignificant.
%

We model the core-envelope coupling following
\citet{Irwin_et_al_07,Gallet_Bouvier_2013} as:
\begin{equation}
    \dot{L}_{coup} =
    \frac{I_{conv}L_{rad}-I_{rad}L_{conv}}{\tau_c(I_{conv}+I_{rad})} -
    \frac{2}{3}R^2_{rad}\omega_{conv}\dot{M}_{rad}
    \label{eq:core_env_coupling}
\end{equation}
where $R_{rad}$ is the radius of the radiative-convective boundary in low-mass stars that parameterizes the coupling torque between the two regions,
$\omega_{conv}$ is the angular velocity of the stellar convective zone,
$M_{rad}$ is the mass of the radiative core of the low-mass stars,
$I_{conv/rad}$ is the moment of inertia of the stellar convective/radiative zone
and $L_{conv/rad}$ is the angular momentum of convective spherical
shell/radiative core of the low-mass star. The first term causes exponential
convergence to solid body rotation on a timescale $\tau_c$, while the second
accounts for the evolution of the core-envelope boundary.

Additionally, the star loses angular momentum due to the stellar winds. We model
that process according to \citet{Schatzman,Irwin_Bouvier_2010,
Gallet_Bouvier_2013,Gallet_Bouvier_15}:

\begin{equation}
    \dot{L}
    =
    -K_w \omega \min(\omega, \omega_{sat})^2
    \sqrt{(R_\star/R_\odot)(M_\odot/M_\star)}
    \label{eq:wind}
\end{equation}

Here $K_w$ is a parameter for the strength of the wind, and $\omega_{sat}$ is
saturation frequency. $R_{\star}$ and $M_{\star}$ are the radius and the mass of the
evolving star respectively.

We treat planets as a single object with solid body rotation, with a fixed mass
($M_{pl}$), radius ($R_{pl}$), and moment of inertia ($I_{pl}=0.26 M_{pl}
R_{pl}^2$).
%
%
    In practice, the angular momentum of the planet is always negligible
    compared to all other angular momenta, and as a result, the planet is always
    held at a pseudo-synchronous spin. That is, the spin rate of the planet very
    quickly adjusts to the value at which the orbit-averaged torque disappears.
    The assumed constant for the moment of inertia comes from Jupiter, but the
    exact value assumed has virtually no effect on the calculated evolution as
    long as it is of that order of magnitude.
%

The evolution defined by the above equations depends on the physical properties
of the star-planet system. Some of those (namely $R_{\star}$, $R_{rad}$,
$M_{rad}$, $I_{conv}$, and $I_{rad}$) change significantly with time. We
approximate that evolution by interpolating among a grid of stellar evolution
tracks calculated using MESA \citep{Paxton_et_al_11, Paxton_et_al_13,
Paxton_et_al_15} in conjunction with MIST \citep{Dotter_16, Choi_et_al_16}.

\subsection{Tidal Dissipation Prescription}
\label{sec:q_prescription}

The two most prominent parameters the tidal phase lag is expected to depend on
based on the range of proposed tidal dissipation theories are the tidal
frequency and the radius of the planet (or its ratio to the radius of a putative
solid core). The radii of the planets in our input sample vary only by about a
factor of two, so we will only include frequency dependence in our prescription
for the phase lag. Furthermore, one can show that to first order in the lag,
$\Delta_{m,s}$ always appears in the tidal equations multiplied by the tidal
Love number $k_2$. As a result, we parameterize
$\Delta_{m,s}'\equiv k_2\Delta_{m,s}$.

The phase lag from section \ref{sec:evolution} is related to the commonly used
tidal quality factor by:
\begin{equation}
    \Delta' = \frac{15}{16\pi Q'}
\end{equation}

We will parameterize the frequency-dependent tidal dissipation on the planet by
a saturating power law:

\begin{equation}
    Q'_{m,s}
    =
    Q_0
    \max\left[
        1,
        \left(\frac{P_{m,s}}{P_{br}}\right)^\alpha
    \right]
    \label{eq:Q_prescription}
\end{equation}
where $P_{m,s}\equiv 2\pi/\tilde{\omega}_{m,s}$, $Q_0$ parameterizes the maximum
dissipation the planet provides, $P_{br}$ parameterizes the tidal period at
which the dissipation saturates, and $\alpha$ parameterizes how quickly the
dissipation decays ($Q'$ grows) as the tidal period increases (decreases) away
from $P_{br}$ for positive (negative) $\alpha$.
%
%
    Note that Eq.
%
%
\ref{eq:Q_prescription}
%
%
    implies that the tidal dissipation of the planet evolves as the orbit
    evolves. More specifically, the orbital period generally gets shorter over
    time, and hence the various $P_{m,s}$ increase.
%

Note that since we will analyze each system separately,
Eq.~\ref{eq:Q_prescription} does not imply a pure power law-dependent
dissipation. Since each system only probes a narrow range of tidal periods, the
power law is only a local approximation. When combining the constraints from all
systems we will treat each tidal period separately, allowing for deviations from
power law behavior.
%
%
    In addition, analyzing one system at a time implies that our prescription
    for $Q'_{m,s}$ need only include dependencies on variables that changed
    during the evolution. In this case, the tidal periods change because the
    orbital period and spin of the planet change so those need to be included.
    However, dependencies on constant quantities (e.g. planet mass) make no sense
    to include since for each individual system that would simply imply a
    different value of $Q_0$. Such dependencies can only be detected when
    combining constraints from multiple systems.
%

In addition, for $\alpha>0$ we tweak Eq.~\ref{eq:Q_prescription} such that for
tidal periods above 20\,d, $Q_{pl}'$ is also held constant. This has the effect
of speeding up calculations dramatically while having only a negligible effect
on the calculated evolution compared to using exactly
Eq.~\ref{eq:Q_prescription}. For more details, see Appendix A of
\citet{penev2022comprehensive}.

Because the dissipation in the star is assumed to have a sub-dominant
effect, we will use a much simpler prescription, namely $Q_\star=const$. This is
entirely appropriate since we only wish to propagate the uncertainty in the
stellar dissipation to the corresponding uncertainty for $Q_{pl}'$. To that end,
we will treat the dissipation in the star as entirely unknown, allowing it to
vary over a very broad range: $5 < \log_{10}Q_\star' < 12$.
%
%
    The final results of this article ultimately are consistent with the
    sub-dominant role of stellar tides, since we find a narrow range of
    $Q_{pl}'$ values that are required to explain the observed circularization. The
    sub-dominance of stellar tides raises the legitimate question of whether it
    is necessary to include the evolution of the stellar structure in our
    calculations instead of assuming a constant stellar radius and moment of
    inertia. However,
%
%
\citet{zahn1989tidal2}
%
%
    show that tidal circularization of late-type main-sequence binary stars is
    often dominated by the pre-main-sequence phase which lasts only a short
    time, but during which the stellar radius is much larger. For that reason, we
    keep the stellar evolution in our calculations.
%

\subsection{Measuring $Q_{pl}'$ from Tidal Circularization}
\label{sec:TidalCircularization}

\begin{figure}
%
    \includegraphics[width=\columnwidth]{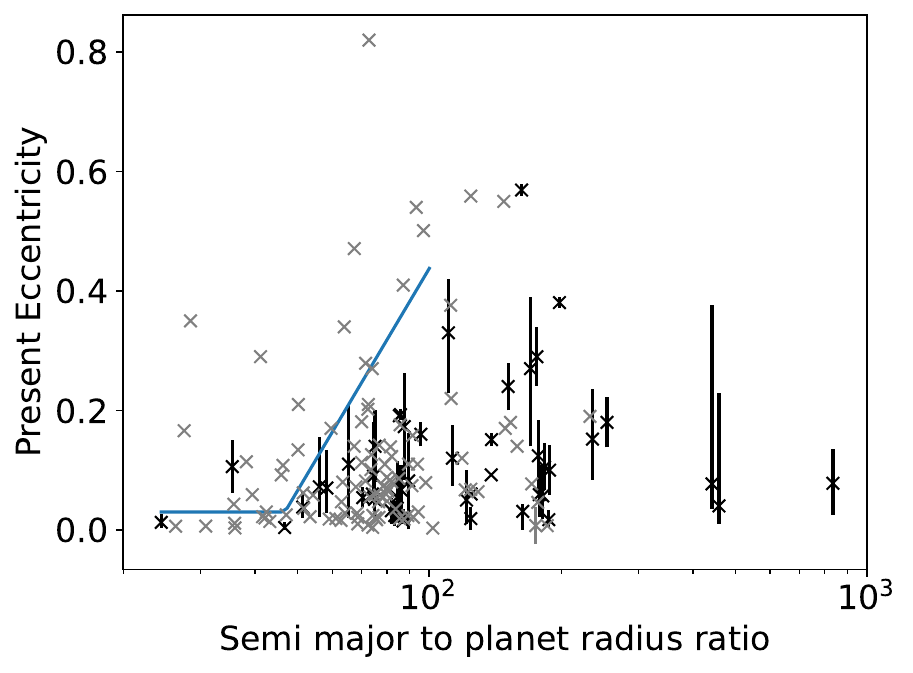}
    \caption{
        Eccentricity vs. semi-major axis to planet radius ratio for the hot
        Jupiter systems selected for analysis: orbital period less than 10\,d,
        planet radius bigger than 0.6 Jupiter radii, stellar mass between 0.4
        and 1.2 $M_{\odot}$.
    }
    \label{fig:e}
\end{figure}

The first step in our procedure is to define an eccentricity envelope: an upper
limit to the eccentricity that a system with a given ratio of the semi-major
axis to planet radius is allowed to have. There must be sufficient tidal
dissipation to ensure that no matter the initial conditions with which tidal
evolution begins, tidal circularization will drive the eccentricity below the
envelope.

Figure \ref{fig:e} shows the scatter plot of present eccentricity vs. semi-major
axis ($a$) divided by planetary radius ($R_{p}$) of different star-exoplanet
systems. The gray crosses correspond to the systems for which only an upper
limit on their present orbital eccentricity is available in the literature. The
black crosses with error bars correspond to those systems for which both upper
and lower uncertainties of the orbital eccentricity are provided. We see a clear
signature of tidal circularization in this plot. When the semi-major axis is
very large then tides are insignificant, and hence, the planetary orbits have a
broad distribution of eccentricities, including very large values. As
$\frac{a}{R_{p}}$ decreases, tides get stronger, and the distribution of
observed eccentricities gradually narrows to smaller and smaller eccentricities.
At even smaller semi-major axes, we see that there are almost no planets with
clearly detected non-zero eccentricities. The evident exception is HAT-P-23 b,
with $a/R_p=35.5$, for which the discovery paper \citep{Bakos_et_al_11} reports
eccentricity of $0.11\pm 0.04$. However, conflicting data is found in
\citet{bonomo2017gaps}, who argue that observations are consistent with a
circular orbit and only find an upper limit on the orbital eccentricity of
0.052. Due to the ambiguity in measured eccentricity, we exclude the HAT-P-23
system from this analysis.

Since the strength of tides is determined by the ratio of the planet size to the
semi-major axis ($R_p/a$), that is the parameter we use for defining the
eccentricity envelope. The solid line in Figure \ref{fig:e} shows the
eccentricity envelope we assume for this analysis, given by:

\begin{equation}
    e_{env}\big(\frac{a}{R_{p}}\big)
    =
    \begin{cases}
        0.03 & \text{if } \frac{a}{R_{p}} <47.8\\
        0.03 + 1.27( \log\frac{a}{R{p}}-1.68) & \text{if }
        47.8<\frac{a}{R_{p}}<130\\
        0.6 & \text{if } \frac{a}{R_{p}}>130
    \end{cases}
    \label{eq:e_env}
\end{equation}

In the presence of tidal dissipation, whatever the initial eccentricity, initial
period, and quality factor $Q$ may be, they must be such that the eccentricity
ultimately becomes its present measured value.  If the tidal dissipation is very
large, then no matter how high the initial eccentricity is, the final calculated
eccentricity at the present age will be below the present measured eccentricity.
Thus, requiring that an initial eccentricity exists which evolves to the
present-day eccentricity sets a maximum for the tidal dissipation, corresponding
to a minimum value of the quality factor. On the other hand, no matter how high
the initial eccentricity is the tidal dissipation should be high enough to
shrink the eccentricity at the present time below the envelope eccentricity.
That is the minimum tidal dissipation, corresponding to the maximum value of the
quality factor.

The procedure for determining if a particular set of tidal dissipation
parameters ($Q_0$, $P_{br}$, $\alpha$) is acceptable, while accounting for the
unknown initial eccentricity, is described as follows. We find an initial
orbital period such that the orbital evolution reproduces the present-day
orbital period at the measured age of the system, starting with an
over-estimated initial eccentricity ($e_i=0.8$) and the tidal dissipation
parameters we wish to evaluate. Because starting with a lower initial
eccentricity would result in a smaller predicted present-day eccentricity, if we
find that the evolution ends with an eccentricity below the measured value for
our system, we can reject the assumed dissipation parameters as too dissipative.
If on the other hand, the final eccentricity is above the envelope, we reject
the dissipation parameters as not dissipative enough.  We accept all parameters
which result in eccentricities between the measured present-day eccentricity and
the envelope eccentricity for the $a/R_p$ of our target hot Jupiter system. We
discuss how we deal with observational uncertainties in Section
\ref{sec:bayesianAnalysis}.

\subsection{Bayesian Analysis}
\label{sec:bayesianAnalysis}

As mentioned in Section \ref{sec:q_prescription}, for each individual hot Jupiter
system, we want to find a joint distribution of $Q_0$, $\alpha$, and $P_{br}$
which is consistent with the observed eccentricity of that particular system
and enforces the envelope eccentricity given the measured $a/R_p$. Given the large number of parameters that must  be specified in order to calculate the
evolution, we are restricted to generating samples from that joint distribution
using a Markov Chain Monte Carlo (MCMC) algorithm. In particular, we will rely on
the \texttt{emcee} python package \citep{foreman2013emcee}, which implements an
ensemble sampling algorithm that uses affine invariant transformation as a
proposal function to simultaneously generate multiple correlated chains (known
as walkers).

The full list of model parameters required to uniquely specify the orbital
evolution is as follows:

\begin{enumerate}
    \item $Q_0$: The minimum tidal quality factor of the planet (maximum
        dissipation) at any frequency (see Eq.~\ref{eq:Q_prescription}).
    \item $\alpha$: The power-law index of the frequency dependence of the
        planetary tidal quality factor (see Eq.~\ref{eq:Q_prescription}).
    \item $P_{br}$: The tidal period at which the planetary dissipation
        saturates (see Eq.~\ref{eq:Q_prescription}).
    \item $Q_\star$: The tidal quality factor of the star.
    \item $M_\star$: mass of the parent star.
    \item $M_p$: mass of the planet.
    \item $\feh$: metallicity of the star.
    \item $t$: Age of the system.
    \item $R_p$: Radius of the planet.
    \item $P_{spin,0}$: Initial spin period of the star.
\end{enumerate}

Standard MCMC terminology splits the distribution we wish to sample from into
priors for the model parameters and a likelihood function ($\mathcal{L}$).  The
priors we wish to impose are determined by the observational constraints
available for the system. For most parameters, we assume independent Gaussian
priors with mean and standard deviation taken directly from Table
\ref{tab:exotable}. However, for transiting planets, the constraints on the
planet's radius come from the transit depth. As a result, the uncertainties in the
planetary and stellar radii are highly correlated. In an effort to account for
this correlation, we use the square of the planet-to-star radius ratio
$(R_p/R_\star)^2$ as a parameter in lieu of the planet radius, whenever
available. In those cases, the planet radius is calculated from the stellar
radius predicted by POET's stellar evolution modeling using the mass, age,
and metallicity of the parent star of a particular sample.

For the tidal parameters and initial stellar spin, we use broad uniform priors:
\begin{eqnarray*}
    \log_{10}Q_0 & \sim & \mathcal{U}[3,12]\\
    \log_{10}P_{br} & \sim & \mathcal{U}[\log_{10}(0.5\,d), \log_{10}(10\,d)]\\
    \alpha & \sim & \mathcal{U}[-5, 5]\\
    \log_{10}Q_\star & \sim & \mathcal{U}[5, 12]\\
    P_{spin,0} & \sim & \mathcal{U}[5\,d, 15\,d]
\end{eqnarray*}
where $\mathcal{U}$ denotes a uniform distribution (in contrast to $U$ from
before which denoted tidal potential).

In addition, several parameters are held fixed for all systems:
\begin{enumerate}
    \item starting age for the tidal evolution: 20\, Myrs
    \item wind strength ($K_w$ in Eq.~\ref{eq:wind}): 0.17\,
        $M_\odot\,R_\odot^2\,\mathrm{day}^2\,\mathrm{rad}^{-2}\,\mathrm{Gyr}^{-1}$

    \item wind saturation frequency ($\omega_{sat}$ in Eq.~\ref{eq:wind}):
        2.78\, rad/day
    \item stellar core-envelope coupling ($\tau_c$ in
        Eq.~\ref{eq:core_env_coupling}): 5 Myr
\end{enumerate}

As we already discussed in section \ref{sec:TidalCircularization}, we need to
define our likelihood function ($\mathcal{L}$) to allow eccentricities anywhere
between the present-day eccentricity of the system and the envelope that
corresponds to the $a/R_p$ value for that system.

The first step is to specify the uncertainty distribution we will assume for the
present-day eccentricity. We begin by noting that eccentricity ($e$) is not
directly constrained by observations, but always in combination with the
argument of periapsis ($\omega$). The complete details of the distribution of
allowed eccentricity values given the observations of a particular system are a
complicated function of the available radial velocity, photometry, and other
observations. It is impractical to attempt to reconstruct these for every single
system, especially given the disparate types, quantity, and quality of
observations available in each case. Instead, we approximate the distribution by
treating $e\cos\omega$ and $e\sin\omega$ as independent parameters, each
following a Gaussian distribution. We further simplify the treatment by assuming
that the standard deviations of the two distributions are the same. These
assumptions imply that the eccentricity follows a Rice distribution. However, we
must be careful with the prior we impose on $e$. By treating $e\cos\omega$ and
$e\sin\omega$ as the variables, we are implicitly imposing a flat prior in that
space. This will have the highly undesirable property that the probability of
$e=0$ is always zero, no matter the available observational data, 
as a Rice distribution is always zero at the origin.
Instead, we wish to impose a uniform prior on $e$. Switching from one prior to
the other simply requires dividing by $e$ and renormalizing, resulting in the
following distribution for the present-day eccentricity up to a normalization
constant:
\begin{equation}
    f(e)
    \propto
    \exp\left(-\frac{e^2+\mu^2}{2\sigma^2}\right)
    I_0\left(\frac{e\mu}{\sigma^2}\right)
\end{equation}
where $\mu$ and $\sigma$ are parameters determining the distribution. They are
related to the means ($\mu_{\cos}$ and $\mu_{\sin}$) and the standard deviation
(assumed the same) of the Gaussian $e\cos\omega$ and $e\sin\omega$
distributions. Namely, $\mu=\sqrt{\mu_{\cos}^2 + \mu_{\sin}^2}$, and $\sigma$ is
the standard deviation (same for both components).

To determine the $\mu$ and $\sigma$ parameters, we consider two cases. When a
two-sided confidence interval is available for the eccentricity (i.e
$e=\bar{e}^{+\delta_+}_{-\delta_-}$ in Table \ref{tab:exotable}), $\mu$ and
$\sigma$ are found by requiring that the 15.9\% and 84.1\% quantiles of the Rice
distribution (before correcting for the eccentricity prior) match the ends of
the confidence interval:

\begin{eqnarray}
    0.159
    =
    \int_{0}^{\bar{e} - \delta_-}
    \frac{e}{\sigma^2} \exp\left(-\frac{e^2+\mu^2}{2\sigma^2}\right)
    I_0\left(\frac{e\mu}{\sigma^2}\right) de\\
    0.841
    =
    \int_{0}^{\bar{e} + \delta_+}
    \frac{e}{\sigma^2} \exp\left(-\frac{e^2+\mu^2}{2\sigma^2}\right)
    I_0\left(\frac{e\mu}{\sigma^2}\right) de
\end{eqnarray}

When observations cannot exclude the possibility of a circular orbit, only an upper limit is available for the eccentricity, we set $\mu=0$ and find $\sigma$ by requiring that the cumulative Rice distribution (before correcting for the eccentricity prior) matches the confidence level of the given limit.

Having defined a distribution for the present-day eccentricity, the next step is
to define the likelihood of our MCMC analysis. Given values for all model
parameters, let $\epsilon(e_i)$ be the function that gives the final
eccentricity the orbital evolution would result in, if started with an initial
eccentricity $e_i$ and orbital period such that the present-day orbital period
is reproduced. Furthermore, let $e_{env}$ be the envelope eccentricity for the
given system, $e_i^{max}$ be the maximum initial eccentricity we consider
(in our case $e_i^{max}=0.8$), and $\pi(e_i)$ be a prior we wish to impose on
the initial eccentricity (the distribution of orbital eccentricities with which
planets begin their tidal evolution). Then up to a normalization constant, the likelihood that the selected model parameters are the true values is:
\begin{equation}
    \mathcal{L}
    \propto
    H\left[\epsilon(e_i^{max}) - e_{env}\right]
    \int_0^{e_i^{max}} f\left[\epsilon(e_i)\right] \pi(e_i) de_i
    \label{eq:likelihood_no_approx}
\end{equation}
where the Heaviside step function $H\left[\epsilon(e_i^{max}) - e_{env}\right]$
imposes the requirement that no initial conditions should result in present day
eccentricity above the envelope, and the integral gives the likelihood of the
system evolving to the observed present-day eccentricity. We do not know the
distribution of initial eccentricities exoplanets start with ($\pi(e_i)$) and
finding $\epsilon(e_i)$ is impractical, so we only evaluate
$\epsilon\left(e_i^{max}\right)$. However, the integral in
Eq.~\ref{eq:likelihood_no_approx} is a weighted average of $f(e)$, so we
approximate the likelihood as:
\begin{equation}
    \mathcal{L}
    \propto
    \begin{cases}
        \frac{1}{e^{max}} \int_0^{e^{max}} f(e) de
        & \text{if } e^{max} \leq e_{env}
        \\
        0 & \text{otherwise}
    \end{cases}
    \label{eq:likelihood}
\end{equation}
where $e^{max}\equiv\epsilon(e_i^{max})$ is the present day eccentricity our
evolution found after starting with $e_i^{max}=0.8$.

\begin{figure}
    \includegraphics[width=1.0\columnwidth]{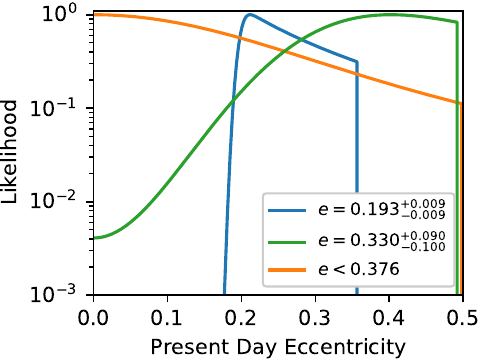}
    \caption{
        Example likelihood functions for a clearly eccentric system, WASP-89 b
        (blue); a system for which a circular orbit is disfavored but not fully
        ruled out, CoRoT-16 b (green); and a system with only an upper limit on the
        eccentricity, HATS-8 b (orange).
    }
    \label{fig:likelihood_comparison}
\end{figure}

Fig.~\ref{fig:likelihood_comparison} shows examples of the likelihood functions
for three different exoplanet systems: WASP-89 b for which a circular orbit is ruled
out to very high significance, CoRoT-16 b for which the available observations
clearly prefer a non-circular orbit but cannot entirely rule out $e=0$, and
HATS-8 b for which only an upper limit is available for the eccentricity.

\begin{figure*}
    \includegraphics[width=1.0\textwidth]{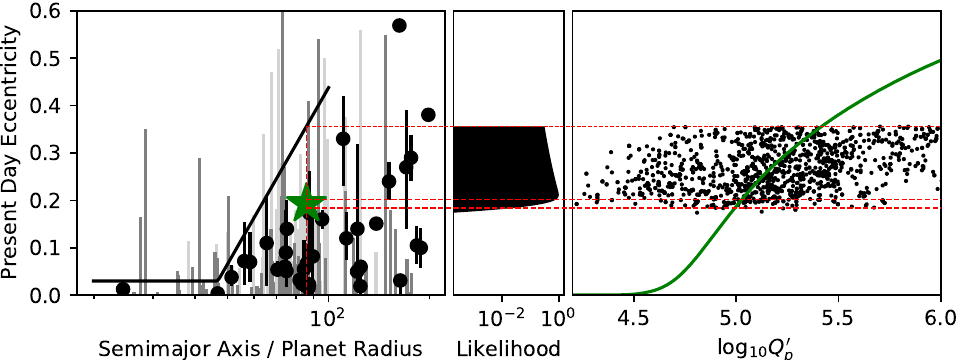}
    \caption{
        Demonstration of the tidal circularization method for constraining
        $Q_{pl}'$. Left: The eccentricity envelope (black line) using the
        observed present-day eccentricities of HJs. Middle: The likelihood
        function, requiring that dissipation is large enough to circularize the
        system to below the envelope no matter how large the initial
        eccentricity is while allowing the observed present-day eccentricity to
        be reproduced.  Right: MCMC samples for $Q_{pl}'(P_{tide}=2\,d)$
        (points), and the relation between predicted present-day eccentricity and
        $Q_{pl}'$ (green curve) for nominal system parameters, and
        $Q_\star'=10^6$. $Q_{pl}'$ was assumed constant only for finding the
        green curve, otherwise, Eq.~\ref{eq:Q_prescription} was
        used.
    }
    \label{fig:circularization_demo}
\end{figure*}

Fig.~\ref{fig:circularization_demo} demonstrates the full MCMC analysis for one
hot Jupiter system: WASP-89 b. The left panel shows the eccentricity envelope and
the observed eccentricity for the system, the middle panel shows the resulting
likelihood function and the right panel shows the MCMC analysis. The green line
shows the final eccentricity as a function of $Q_{pl}'$ assuming a constant
$Q_{pl}'$ model (i.e. $\alpha=0$ in Eq.~\ref{eq:Q_prescription}),
$Q_\star'=10^6$ and fixing all system properties to their nominal values. The
points correspond to the MCMC samples of frequency-dependent $Q_{pl}'$ (per
Eq.~\ref{eq:Q_prescription}). For each MCMC sample, we evaluated
Eq.~\ref{eq:Q_prescription} at a tidal period of 2\, days, which is the tidal
period this particular system is most sensitive to, i.e. where that particular
system gives the tightest constraints on $Q_{pl}'$.

\subsection{Constraining Frequency Dependent $Q_{pl}'$ and Convergence Test}
\label{sec:constraints_convergence}

A particularly useful approach for building constraints for the value of
$Q_{pl}'$ at each tidal period begins by evaluating Eq.~\ref{eq:Q_prescription}
for each MCMC sample of a given system on a grid of tidal periods, obtaining
samples of $Q_{pl}'(P_{tide}, R_{pl})$ at each grid point. After that, again at
each of the grid points, we create a kernel density estimate of the probability
distribution of $Q_{pl}'(P_{tide})$ for each system. Finally, the probability
distributions of all systems at a given grid point are multiplied together and
re-normalized to get a combined constraint. Since each individual system probes
a relatively narrow range of tidal periods, the combined constraint produced in
this fashion is no longer restricted to a purely power-law dependence of
$Q_{pl}'$, but rather Eq.~\ref{eq:Q_prescription} simply ensures continuity and
our prior on $\alpha$ imposes a limit on the maximum slope of the dependence of
$\log_{10}Q_{pl}'$ on $\log_{10}P_{tide}$. When combining constraints, we apply
one additional consideration. Because each individual binary is sensitive only
to the dissipation in a relatively narrow range of periods, it produces tight
constraints on $Q_{pl}'(P_{tide})$ only in that range. Moving away from that
period, the spread in allowed values of $Q_{pl}'(P_{tide})$ is limited by the
prior we impose on $\alpha$. When building the combined constraints, for each
binary we identify the range of periods where the constraints are tightest and
only include that binary in the combined constraint for the $P_{tide}$ grid
points within that range.

MCMC sampling using the emcee \citep{foreman2013emcee} package involves
advancing multiple (in our case 64) chains that are not statistically
independent. Since we don't know \textit{a priori} how the likelihood function looks,
we initialize the chains with samples that are over-dispersed compared to the
equilibrium posterior distribution. The chains gradually migrate to
higher likelihood regions until the equilibrium distribution is found. As a
result, early samples tend to overestimate the uncertainties, and possibly
introduce bias in estimated values. The usual practice is to discard these so-called `burn-in' initial samples. We thus need a method to estimate this
burn-in period for our samples. Subsequently, we need an estimate
of how many additional iterations need to be accumulated to get
reliable estimates of the desired quantities and their uncertainty
distributions.

In this work, we are aiming to measure the quantiles of the distribution of
$Q_{pl}'$ at a grid of tidal periods. The method of determining the number of
burn-in steps required to reach close to the equilibrium distribution, 
and the number of steps required after the burn-in to measure the value of the
cumulative distribution function at each quantile with a desired precision is
discussed in \citet{patel2022constraining} and \citet{penev2022comprehensive}.
The method developed in these papers is an extension of the method discussed in
\citet{raftery1991many}, adapting it for use with multiple interdependent
chains. We have applied the same method discussed in these papers to perform a
convergence analysis of the MCMC process in this project.

Essentially, for each quantile ($q$) the algorithm finds a simultaneous burn-in
period ($N_b$) and quantile values that meet the following two conditions:
\begin{enumerate}
    \item The distribution of the fraction of the 64 $N_b+1$ samples smaller
        than $q$ is within $10^{-3}$ of the distribution after infinitely many
        steps.
    \item The fraction of all samples after $N_b$ below $q$ is equal to the
        target quantile.
\end{enumerate}
That same algorithm also provides an estimate of the standard deviation of the
fraction of samples below $q$ over different realizations of a chain of a given
length. This is a measure of how precisely a given quantile is estimated.

\section{Results}
\label{sec:results}

\subsection{Individual Constraints}
\label{sec:individual_constraints}

\begin{figure*}
    \includegraphics[width=0.95\textwidth]{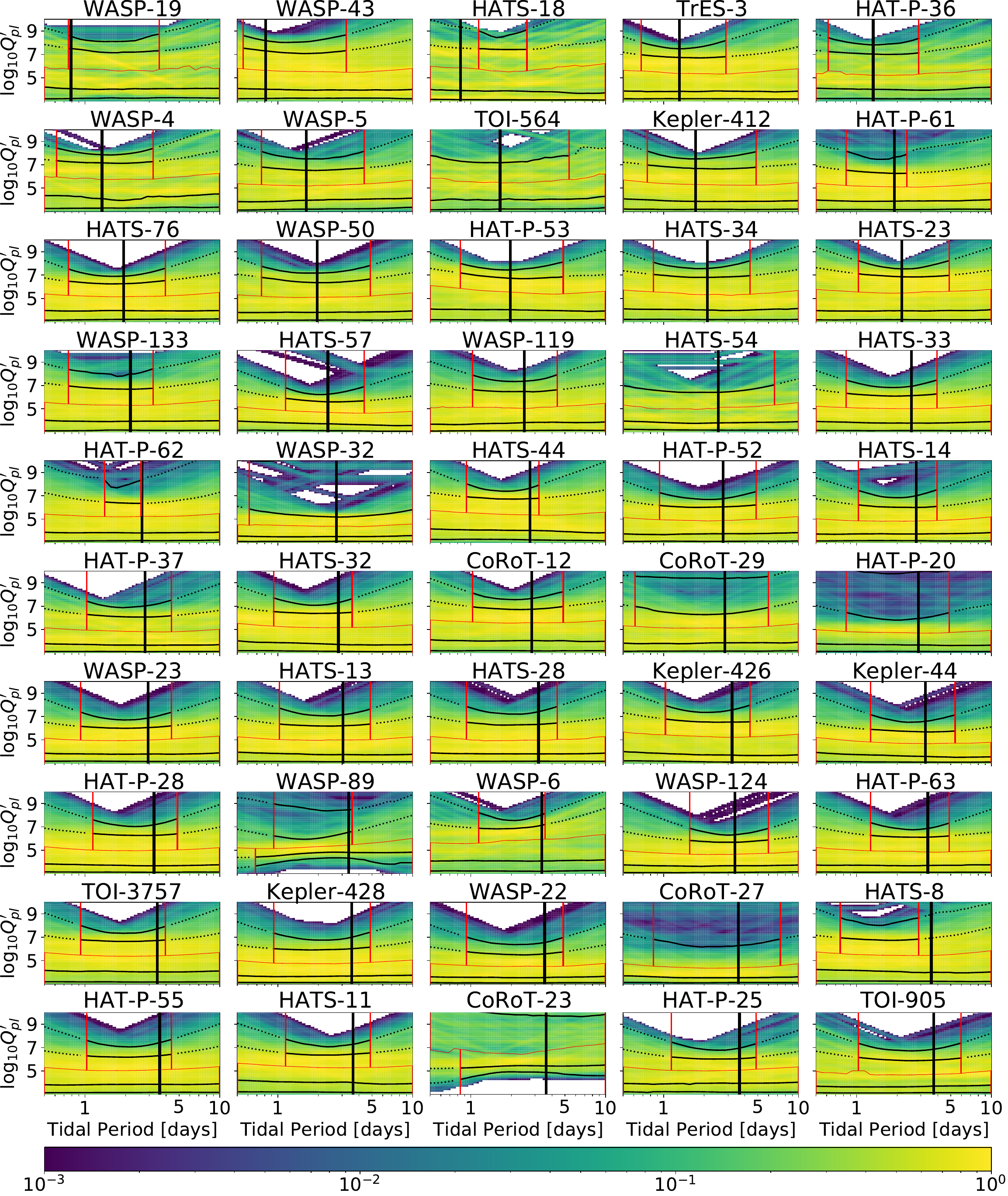}
    \caption{
        Constraints on $\log_{10}Q'_{pl}$ for tidal periods in the range $0.5\,d
        < P_{tide} < 10\,d$ for each individual star-exoplanet system.  The
        color indicates the probability density of $\log_{10}Q'_{pl}$ at each
        tidal period. The thick black vertical line shows the system's orbital
        period. The left-to-right red curve shows the median of the
        $\log_{10}Q'_{pl}(P_{tide})$ distribution at each tidal period.
        Similarly, the four black curves from bottom to top represent the 2.3,
        15.9, 84.1, and 97.7 percentiles, respectively, of the distribution of
        $\log_{10}Q'_{pl}$ at each tidal period. In the region between the two
        vertical red lines, we consider the constraints to be driven by the
        available measurements for that system, instead of by the prior we place
        on the period dependence power-law index $\alpha$ (see text for
        details).
    }
    \label{fig:indivCons1}
\end{figure*}

\begin{figure*}
    \includegraphics[width=0.95\textwidth]{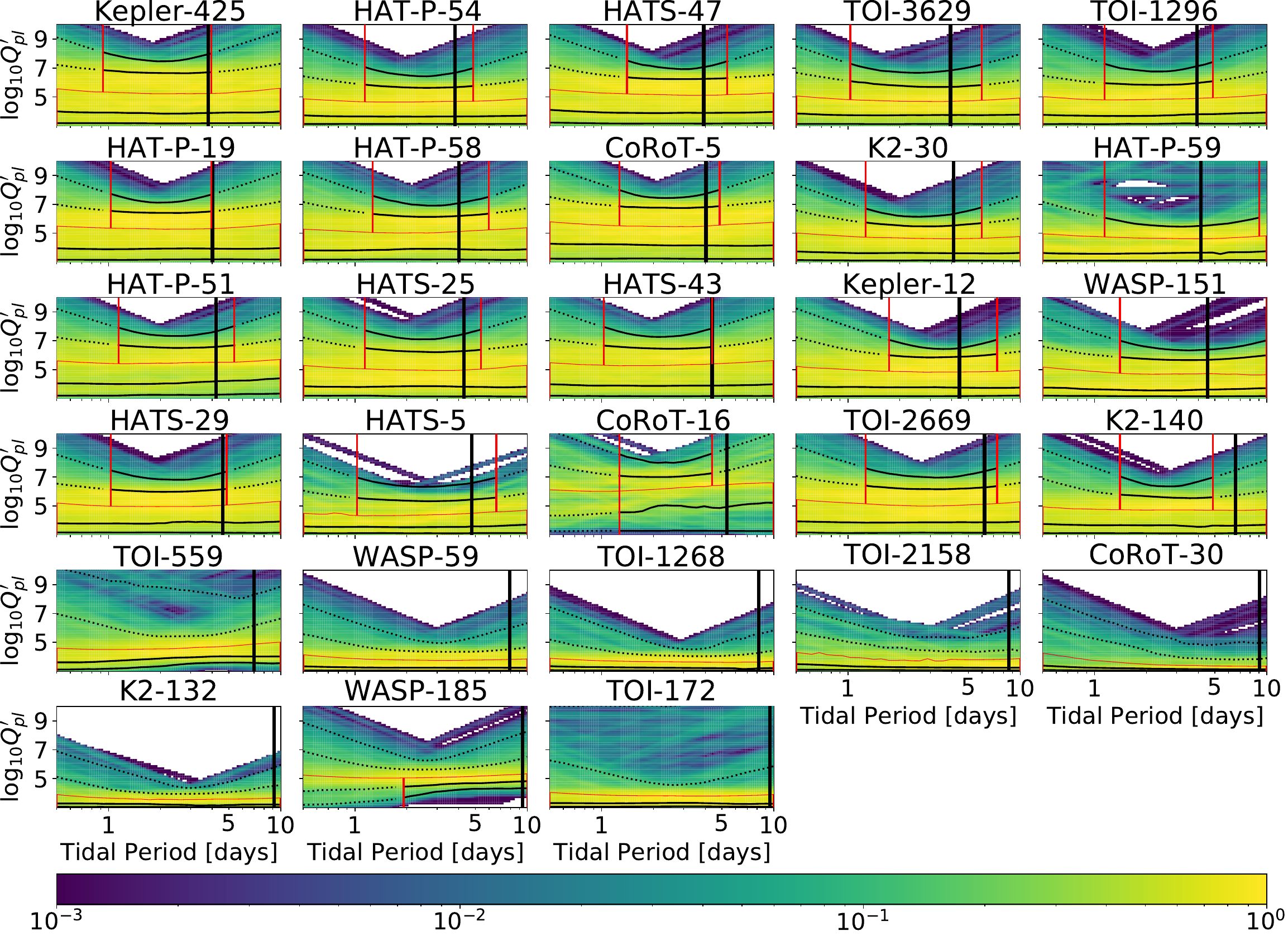}
    \caption{Continuation of Figure \ref{fig:indivCons1}.}
    \label{fig:indivCons2}
\end{figure*}

Figures \ref{fig:indivCons1} and \ref{fig:indivCons2} show the constraints
on $\log_{10}Q'_{pl}$ for different tidal periods, $P_{tide} \in [0.5, 10]$ days
for each individual star-exoplanet system. For every sample of the MCMC, the
values of the parameters ($\alpha$, $P_{br}$ and $Q_0$) are used to calculate
$Q_{pl}'(P_{tide})$ using Eq.~\ref{eq:Q_prescription} at a grid of tidal
periods. Thus, for each system at every value of $P_{tide}$, we get samples of
$\log_{10}Q'_{pl}$ required by the observed and envelope eccentricities of the
system. Using Kernel Density Estimation (KDE) with an Epanechnikov kernel
function with a width of 0.2 dex, we construct an estimate of the probability
density function (PDF) of $\log_{10}Q'_{pl}(P_{tide})$ at each tidal period.
The color along each vertical slice of a panel in Fig.~\ref{fig:indivCons1} and
Fig.~\ref{fig:indivCons2} shows that PDF at the tidal period of the slice. The
thick black vertical line shows the orbital period of the system. The thin red
curve going from left to right shows the median of that PDF at each tidal
period, and the black curves from the bottom to top show the 2.3, 15.9, 84.1,
and 97.7 percentiles, respectively. Only samples after the automatically
determined burn-in period were used to generate Fig.~\ref{fig:indivCons1} and
\ref{fig:indivCons2}.

A common feature seen in many of the individual constraints is that for a given
system, $Q_{pl}'$ is only constrained for a narrow range of tidal periods, and
moving away from that, the distribution of allowed $Q_{pl}'$ values gets much
broader. This is caused by the fact that for an individual system, tidal
circularization is dominated by a narrow range of tidal periods. As we plan to
use the individual constraints to build a combined constraint, for each system
we need to define the range of tidal periods that the system is sensitive to (which we dub the reliable period range) and only include its constraints at those tidal
periods. In fact, for each system, we define two reliable period ranges -- one
for the upper half and one for the lower half of the distribution. These ranges
are defined using a very simple criterion. Namely, for the upper/lower half of
the distribution, the range is defined by requiring that the 97.7\%/2.3\%
percentile is within 0.7 dex of its minimum/maximum value. The reliable period
ranges are delineated by the vertical red lines in each plot (one set for the
distribution above the median and another set for the distribution below).

Figures \ref{fig:burn1} and \ref{fig:burn2} show the estimated burn-in period.
The solid
portions of the curves correspond to the reliable period range for each quantile
where the results are dominated by observational data. The dotted portions of
the curves are outside the reliable period range. The black region depicts the
total number of emcee steps accumulated for that system. As these figures
demonstrate, for all systems we have generated sufficient samples to exceed the
burn-in period for each quantile of $Q_{pl}'$ for each tidal period within the
reliable period range.

Figures \ref{fig:sigmacdf1} and \ref{fig:sigmacdf2} show the standard deviation
of the value of the cumulative density function at the estimated value of the four
percentiles ($\sigma_{CDF}$) for different values of $P_{tide}$ for the steps of
MCMC, generated in the post-burn-in epoch. We see that for all systems,
$\sigma_{CDF} << \min(CDF, 1-CDF)$ for each quantile. In other words, the
fraction of samples outside a given confidence interval is estimated with high
precision (small uncertainty). For example, for WASP-19, the fraction of samples
below the 15.9\% percentile is within 1\% or better of 15.9\% for all tidal
periods.


Figures \ref{fig:indivCons1} and \ref{fig:indivCons2} show that for most of
the star-exoplanet systems, we get a sharp upper limit for $\log_{10} Q'_{pl}$ for
a range of values of $P_{tide}$. If the value of $\log_{10} Q'_{pl}$ is greater
than this upper limit then the tidal dissipation is not strong enough to shrink
the orbital eccentricity below the envelope eccentricity for the system in
the scatter plot of eccentricity vs. $\log(\frac{a}{R_{p}})$. On the other hand,
except for WASP-89 b, CoRoT-23 b, and WASP-185 b, the lower limits are less
sharp, though for many systems the likelihood drops off substantially for low
$Q_{pl}'$ values (note that the color scale in the plots is logarithmic).

\begin{figure*}
    \includegraphics[width=0.85\textwidth]{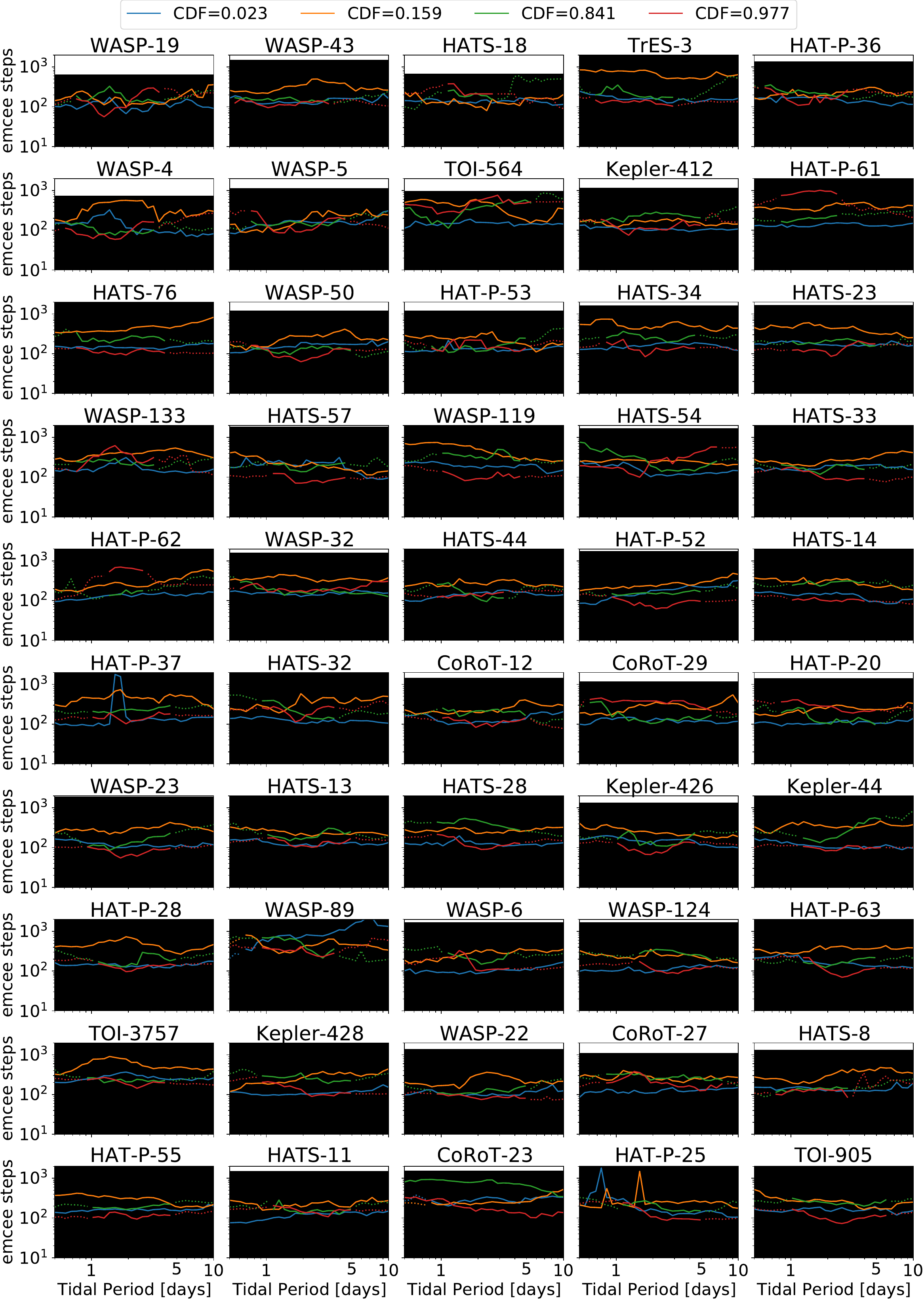}
    \caption{
        Estimated emcee steps required to reach close to the equilibrium value
        of the percentiles of $\log_{10}Q'_{pl}$ as a function of $P_{tide}$.
        Within the solid portion of the curves, constraints are influenced by the
        observation and hence are reliable. The dotted portion of the curves is
        influenced by our prior assumptions. The black area represents the
        number of steps completed.
    }
    \label{fig:burn1}
\end{figure*}

\begin{figure*}
    \includegraphics[width=0.85\textwidth]{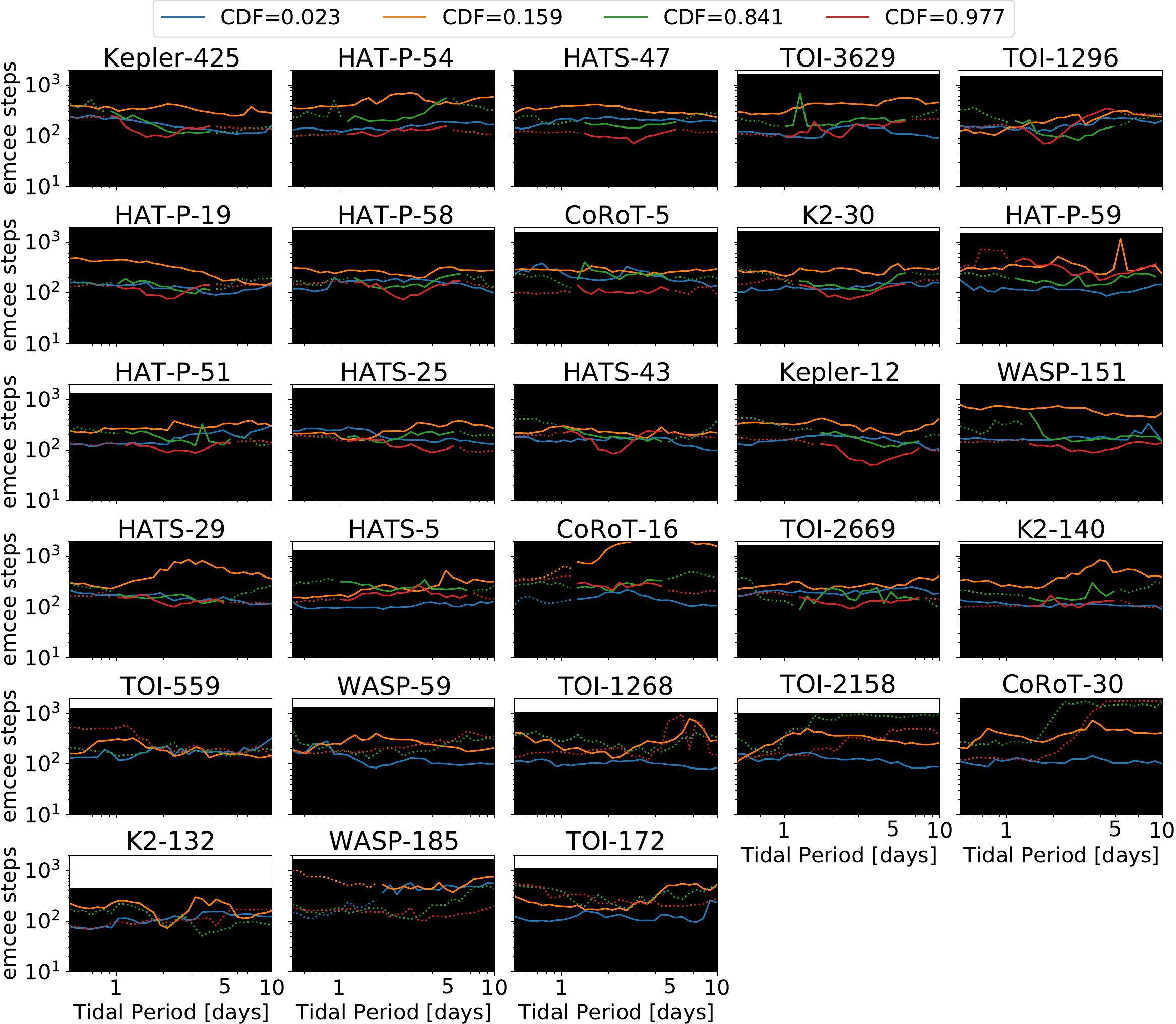}
    \caption{Continuation of Figure \ref{fig:burn1}.}
    \label{fig:burn2}
\end{figure*}

\begin{figure*}
    \includegraphics[width=0.85\textwidth]{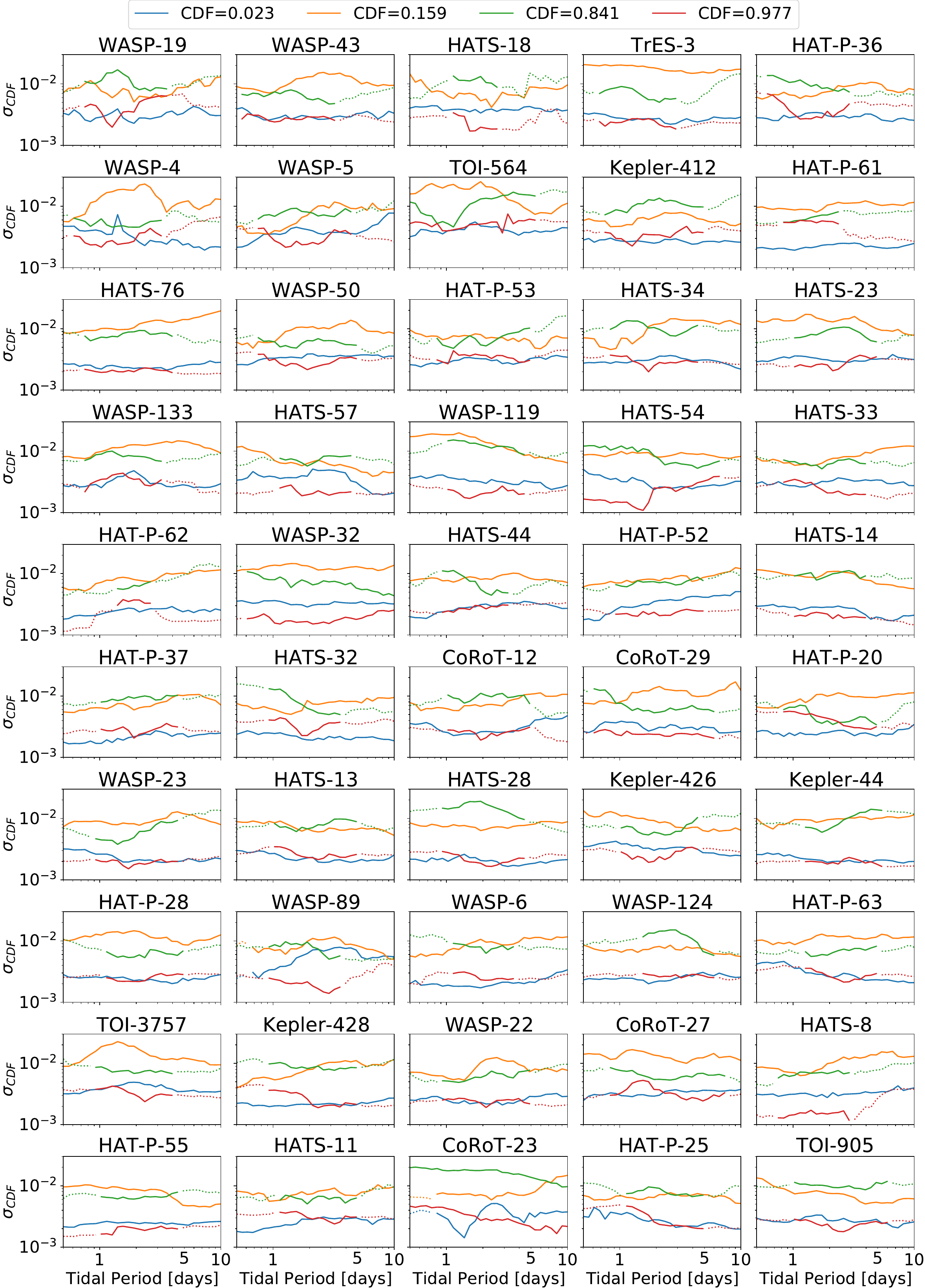}
    \caption{
        The standard deviation of the cumulative density function of
        $\log_{10}Q'_{pl}$ up to the indicated percentile as a function of
        $P_{tide}$. Within the solid portion of the curves, constraints are
        influenced by the observations and hence are reliable. The dotted portion of
        the curves is influenced by our prior assumptions.
    }
    \label{fig:sigmacdf1}
\end{figure*}

\begin{figure*}
    \includegraphics[width=0.85\textwidth]{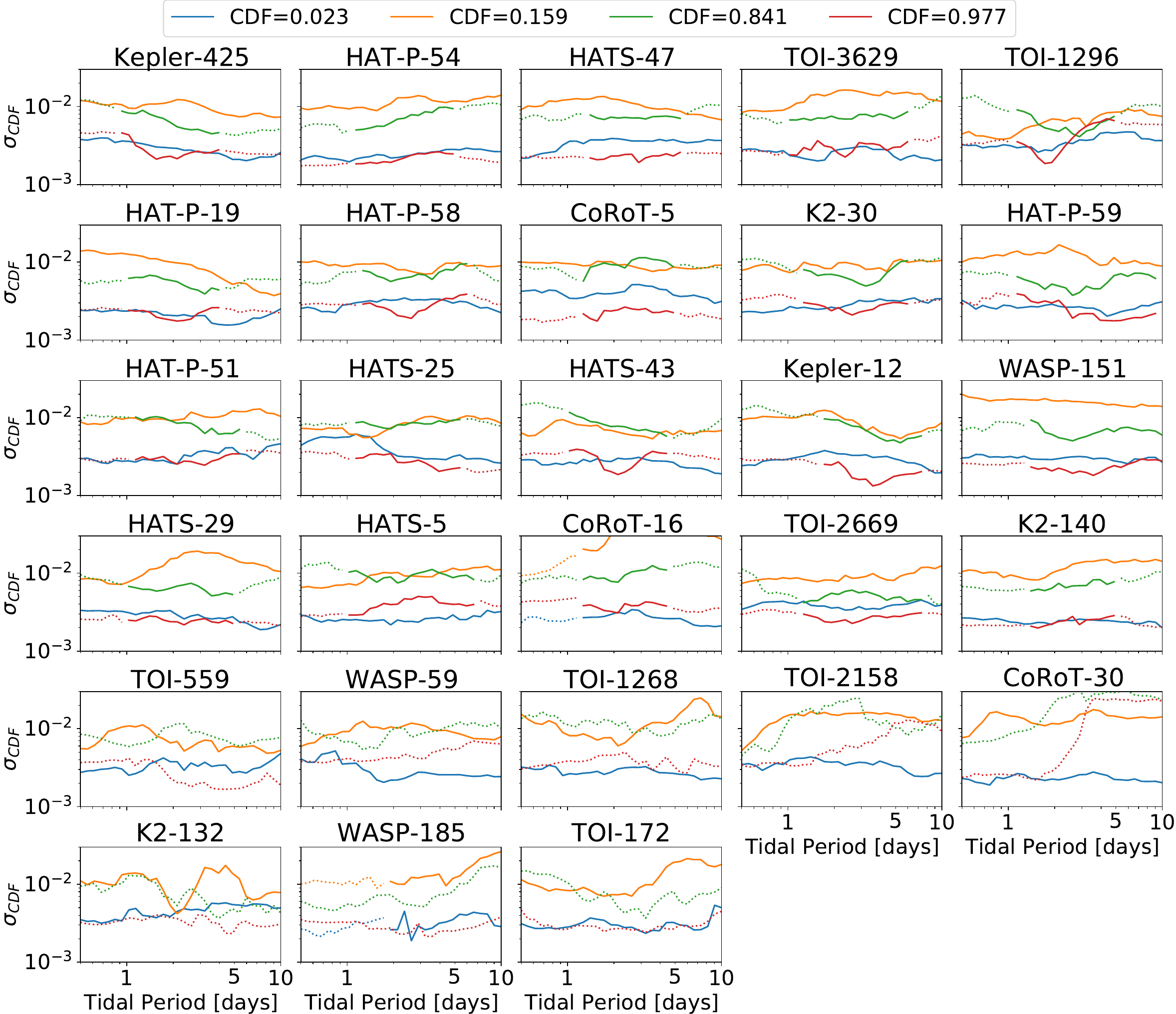}
    \caption{Continuation of Figure \ref{fig:sigmacdf1}.}
    \label{fig:sigmacdf2}
\end{figure*}

For CoRoT-23 b, our analysis shows lower limits of $\log_{10}Q'_{pl}$ for a
range of values of $P_{tide}$. Had the $\log_{10}Q'_{pl}$ been less than the
lower limit for any value of $P_{tide}$ within that range, the tidal dissipation
would become so strong that the orbital eccentricity of the system at present
age would have been smaller than the actual orbital eccentricity the system has
now. We require that the tidal dissipation not be that strong, as the system
started the evolution with a very high orbital eccentricity of 0.8. However, as
mentioned in Section \ref{sec:data}, the significantly non-zero value of the
eccentricity we used for the analysis (reported by \citet{rouan2012transiting})
has been disputed \citep{bonomo2017gaps}. As a result, we show the individual
constraints for CoRoT-23 b but do not include them when constructing the
combined constraints on $\log_{10}Q'_{pl}$.

\subsection{Combined Constraints}
\label{sec:combined_constraints}

\begin{figure}
    \includegraphics[width=\columnwidth]{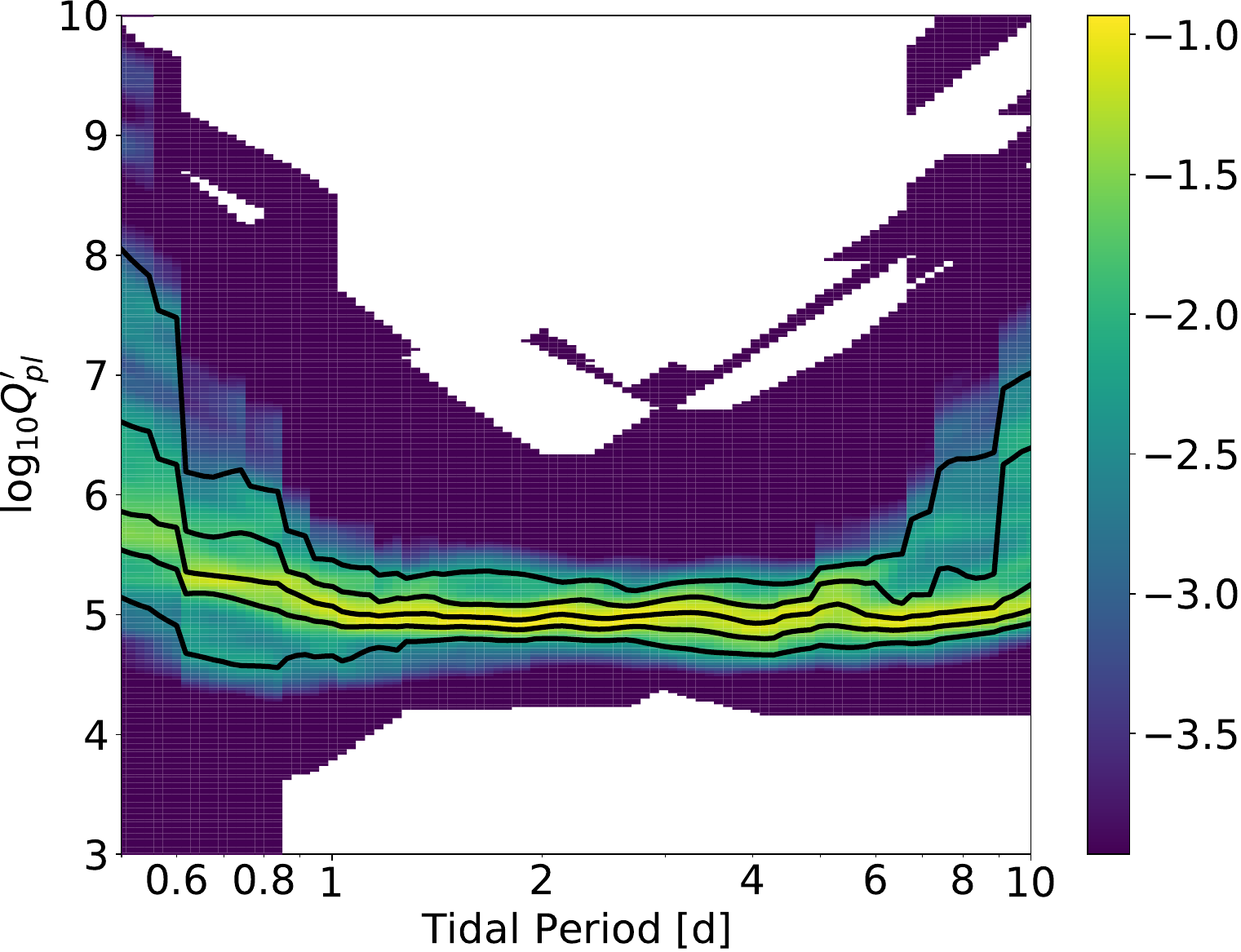}
    \caption{
        Combined constraints for $\log_{10}Q'_{pl}$ assuming a common frequency
        dependent dissipation applies to all planets from Table
        \ref{tab:exotable} except CoRoT-23 b. The color shows $\log_{10}$
        of the combined probability. The five black curves from the bottom
        to top show the 2.3\%, 15.9\%, 50.0\%, 84.1\%, and 97.7\% percentiles of
        the $\log_{10}Q'_{pl}$ distribution at each tidal period.  For every
        $P_{tide}$, the combined posterior probability density distribution of
        $\log_{10}Q'_{pl}$ is the product of the posterior probability density
        distribution of $\log_{10}Q'_{pl}$ for individual systems for which that
        tidal period falls within the reliable period range.
    }
    \label{fig:comb}
\end{figure}

Next, we combine the constraints on $\log_{10}Q'_{pl}$ for individual exoplanet
systems to build constraints for the tidal quality factor under the assumption
that the planets in our sample are similar enough to share the same tidal
dissipation. In order to do so, for each tidal period ($P_{tide}$), we multiply
together the individual probability density functions for
$\log_{10}Q'_{pl}(P_{tide})$. The product only includes posterior probability
density functions of $\log_{10}Q'_{pl}$ for systems for which a given tidal
period falls within the reliable period range. If a tidal period falls in the
range of the tidal periods of a particular system for which only upper/lower
percentiles above/below the median are reliable then we replace the unreliable
lower/upper portion of the probability density of $\log_{10}Q'_{pl}$ with the
probability density of its median at that tidal period and re-normalize the
distribution before including it. The resulting product of PDFs is then
normalized to unit integral.

When building the combined constraints we exclude 
three
systems: CoRoT-23 b, K2-132 b, and TOI-2669 b. CoRoT-23 b was excluded because
its eccentricity measurement is suspect, as explained before. K2-132 b and
TOI-2669 b were excluded because their host stars have evolved to their red giant
phase. For such ages, the POET handling of stellar evolution is potentially
suspect as it involves interpolating among fast-evolving post-main-sequence
portions of the stellar evolution tracks.

In addition, per Eq.
\ref{eq:e_env}
%
%
    we imposed an envelope eccentricity of 0.6 for systems with $a/R_p>130$
    which is not physically justified.  Conservative assumption would suggest
    that for those systems there is no observable tidal circularization. To
    account for that, we exclude the upper limit on $\log_{10}Q'_{pl}$ for these
    systems: CoRoT-30 b, WASP-59 b, WASP-185 b, TOI-559 b, K2-132 b, TOI-2158 b,
    TOI-172 b, TOI-1268 b. In practice, this means that from these systems,
    only WASP-18 b contributes to the combined constraint, giving a lower limit
    to $Q_{pl}'$ at the longest tidal periods studied.
%

Figure \ref{fig:comb} shows the resulting combined constraint for
$\log_{10}Q'_{pl}$ for the period range $0.5\,d < P_{tide} < 10\,d$. The
color shows $\log_{10}$ of the combined probability, and the black curves,
from bottom to top, represent 2.3\%, 15.9\%, 50\%, 84.1\%, and 97.7\%
percentiles of the $\log_{10}Q'_{pl}$ distribution at each tidal period,
respectively.

\begin{figure}
    \includegraphics[width=\columnwidth]{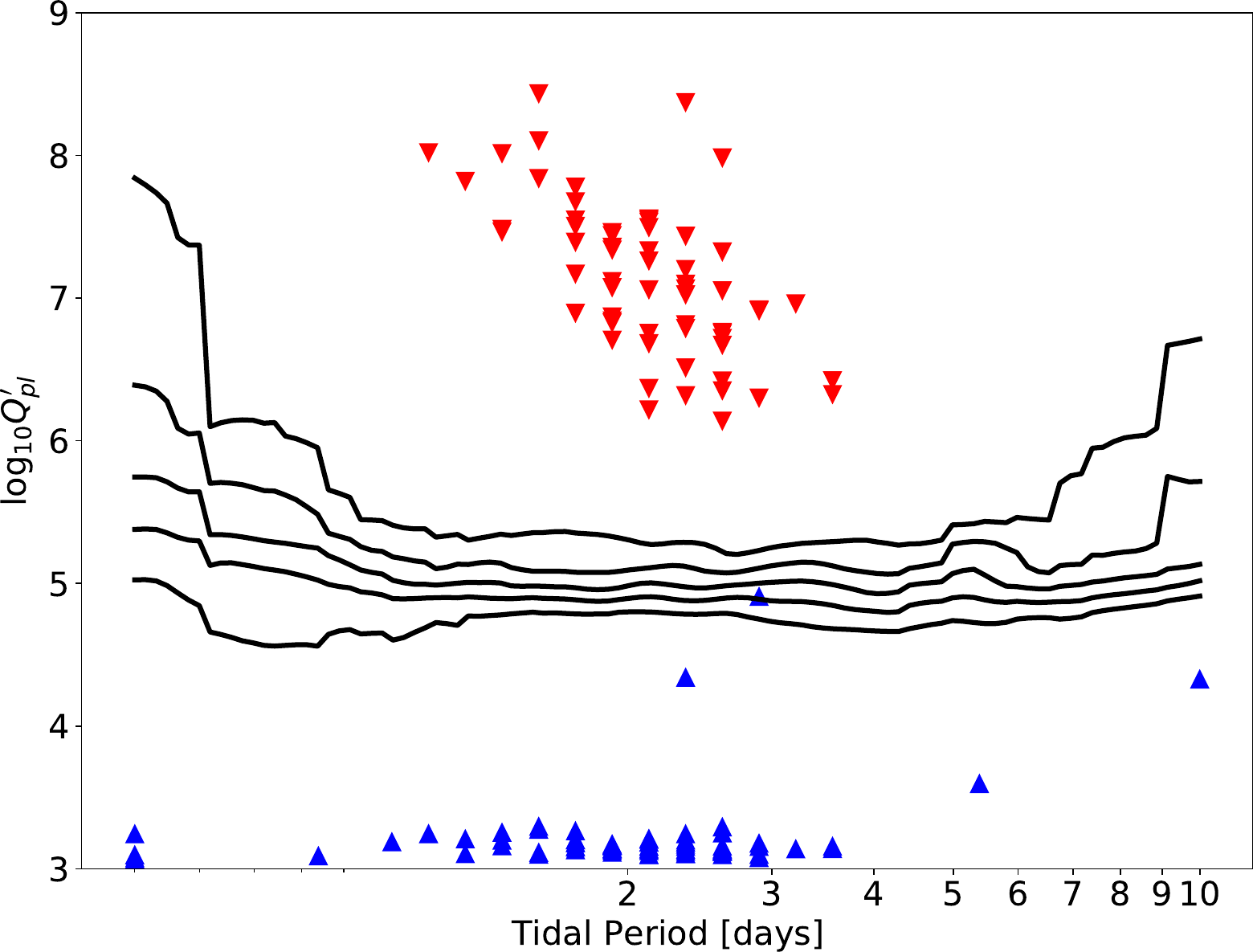}
    \caption{
        Black curves: 2.3\%, 15.9\%, 50.0\%, 84.1\%, and 97.7\% percentiles of
        the $\log_{10}Q'_{pl}$, as in Fig.~\ref{fig:comb}.
        Red (blue) triangles pointing downward (upward) show the 97.7\% (2.3\%)
        percentile of the constraint for an individual system at the tidal period where for that particular system the quantile is the lowest
        (highest).
    }
    \label{fig:indivVsComb}
\end{figure}

The combined constraint is only meaningful if the individual constraints are
actually consistent with each other. Fig.~\ref{fig:indivVsComb} compares the
combined constraints with the constraints for individual systems. The five black
curves are as described in the previous paragraph for Fig.~\ref{fig:comb}.
For each system,
a red and a blue triangle are shown, indicating the minimum of the 97.7\% and
the maximum of the 2.3\% percentiles for that system's constraints.

%
    To see why the combined constraint is significantly tighter than even the tightest individual constraints, consider for example what would be required for $Q_{pl}$ to be near the lower limit for individual systems. Ignoring for a moment the uncertainties in anything other than present-day eccentricity, the lower limit on $Q_{pl}$ comes from starting the evolution with an initial eccentricity of $e_i=0.8$ and landing near the lower limit of the observationally determined distribution of the present-day eccentricity. Thus, this would imply that all literature eccentricities are systematically 2-$\sigma$ below the implied maximum likelihood value. More realistically, the true values of the present-day eccentricity, stellar mass, and stellar age must be simultaneously below the maximum likelihood values reported by about 1-$\sigma$, while the planet's mass and radius must be above by about 1-$\sigma$ systematically for all exoplanet systems. Similarly, the 97th percentile of $Q_{pl}$ for each individual system comes from assuming systematically larger age and planet radius, combined with systematically smaller stellar mass.
%

As Fig. \ref{fig:indivVsComb} shows, the combined constraints are consistent
with all individual constraints, even for the systems that were excluded from
the combined constraint due to suspicion that the stellar interpolation may not
be reliable or for the one with disputed eccentricity measurement. Table
\ref{tab:Comp} shows the comparison in tabular form.

We see that, for $P_{tide}$ between 0.8 and 7 days, the combined constraints on
$\log_{10}Q'_{pl}$ restrict its value to a very narrow range, and apart from the
two exceptions are consistent with all individual constraints.
We also see that the black curves are approximately parallel to the x-axis,
implying that $\log_{10}Q'_{pl}$ is approximately independent of the tidal
period, and over that period range, typical hot Jupiter planets have
$\log_{10}Q'_{pl}=5.0\pm0.5$.

\begin{table*}
    {
        \centering
        \caption{
            Comparison of combined and individual percentiles of
            $\log_{10}Q'_{pl}$ for the $P_{tide}$ values at which each system
            provides the tightest constraints.
        }
        \label{tab:Comp}
        \resizebox{0.70\textwidth}{!}{
            \begin{tabular}{c@{\quad\quad\quad}c@{\quad\quad\quad}ccccc@{\quad\quad\quad}ccccc}
    &&\multicolumn{5}{c@{\quad\quad\quad}}{\textbf{Individual Quantiles of $\log_{10}Q'_{p}$}} & \multicolumn{5}{c}{\textbf{Combined Quantiles of $\log_{10}Q'_{p}$}}\\
\multicolumn{1}{c@{\quad\quad\quad}}{\textbf{System}} & \multicolumn{1}{c@{\quad\quad\quad}}{$\mathbf{P_{tide}}$} & \multicolumn{1}{c@{}}{\textbf{2.3\%}} & \multicolumn{1}{c@{}}{\textbf{15.9\%}} & \multicolumn{1}{c@{}}{\textbf{50.0\%}} & \multicolumn{1}{c@{}}{\textbf{84.1\%}} & \multicolumn{1}{c@{\quad\quad\quad}}{\textbf{97.7\%}} & \multicolumn{1}{c@{}}{\textbf{ 2.3\%}} & \multicolumn{1}{c@{}}{\textbf{ 15.9\%}} & \multicolumn{1}{c@{}}{\textbf{ 50.0\%}} & \multicolumn{1}{c@{}}{\textbf{ 84.1\%}} & \multicolumn{1}{c@{}}{\textbf{ 97.7\%}} \\
\hline
\hline
WASP-19 b & 1.56 & 3.27 & 4.03 & 5.63 & 7.14 & 8.10 & 4.79 & 4.90 & 4.98 & 5.09 & 5.36 \\
WASP-43 b & 1.14 & 3.24 & 3.97 & 5.55 & 7.14 & 8.02 & 4.70 & 4.90 & 4.99 & 5.14 & 5.36 \\
HATS-18 b & 1.56 & 3.11 & 3.87 & 5.58 & 7.43 & 8.43 & 4.79 & 4.90 & 4.98 & 5.09 & 5.36 \\
TrES-3 b & 1.40 & 3.20 & 3.84 & 5.36 & 6.72 & 7.48 & 4.78 & 4.89 & 5.00 & 5.14 & 5.34 \\
HAT-P-36 b & 1.27 & 3.21 & 4.01 & 5.22 & 7.03 & 7.82 & 4.75 & 4.90 & 5.01 & 5.13 & 5.31 \\
WASP-4 b & 1.56 & 3.30 & 4.00 & 5.74 & 7.14 & 7.84 & 4.79 & 4.90 & 4.98 & 5.09 & 5.36 \\
WASP-5 b & 2.12 & 3.20 & 3.98 & 5.23 & 6.52 & 7.50 & 4.80 & 4.91 & 5.00 & 5.11 & 5.27 \\
TOI-564 b & 1.03 & 3.19 & 4.00 & 5.54 & 7.22 & 10.78 & 4.61 & 4.90 & 5.03 & 5.19 & 5.41 \\
Kepler-412 b & 1.73 & 3.18 & 3.86 & 5.31 & 6.67 & 7.51 & 4.79 & 4.89 & 4.97 & 5.08 & 5.35 \\
HAT-P-61 b & 1.40 & 3.16 & 3.87 & 5.14 & 6.32 & 7.47 & 4.78 & 4.89 & 5.00 & 5.14 & 5.34 \\
HATS-76 b & 1.73 & 3.20 & 3.97 & 5.14 & 6.25 & 6.90 & 4.79 & 4.89 & 4.97 & 5.08 & 5.35 \\
WASP-50 b & 1.73 & 3.15 & 3.80 & 5.10 & 6.37 & 7.17 & 4.79 & 4.89 & 4.97 & 5.08 & 5.35 \\
HAT-P-53 b & 1.92 & 3.16 & 4.04 & 5.55 & 6.73 & 7.44 & 4.79 & 4.88 & 4.96 & 5.08 & 5.33 \\
HATS-34 b & 1.73 & 3.20 & 4.02 & 5.33 & 6.80 & 7.55 & 4.79 & 4.89 & 4.97 & 5.08 & 5.35 \\
HATS-23 b & 2.12 & 3.18 & 3.95 & 5.46 & 6.86 & 7.53 & 4.80 & 4.91 & 5.00 & 5.11 & 5.27 \\
WASP-133 b & 1.73 & 3.18 & 3.92 & 5.32 & 6.69 & 7.78 & 4.79 & 4.89 & 4.97 & 5.08 & 5.35 \\
HATS-57 b & 2.12 & 3.14 & 3.72 & 4.71 & 5.62 & 6.22 & 4.80 & 4.91 & 5.00 & 5.11 & 5.27 \\
WASP-119 b & 2.12 & 3.14 & 3.78 & 5.06 & 6.47 & 7.33 & 4.80 & 4.91 & 5.00 & 5.11 & 5.27 \\
HATS-54 b & 0.50 & 3.24 & 3.95 & 5.33 & 6.98 & 11.77 & 5.03 & 5.38 & 5.75 & 6.39 & 7.84 \\
HATS-33 b & 1.92 & 3.14 & 3.78 & 5.02 & 6.09 & 6.83 & 4.79 & 4.88 & 4.96 & 5.08 & 5.33 \\
HAT-P-62 b & 1.73 & 3.16 & 3.85 & 5.21 & 6.40 & 7.68 & 4.79 & 4.89 & 4.97 & 5.08 & 5.35 \\
WASP-32 b & 0.84 & 3.09 & 3.50 & 4.51 & 5.61 & 9.96 & 4.57 & 5.02 & 5.24 & 5.48 & 5.93 \\
HATS-44 b & 1.73 & 3.26 & 4.05 & 5.41 & 6.73 & 7.39 & 4.79 & 4.89 & 4.97 & 5.08 & 5.35 \\
HAT-P-52 b & 2.12 & 3.10 & 3.75 & 4.84 & 6.02 & 6.68 & 4.80 & 4.91 & 5.00 & 5.11 & 5.27 \\
HATS-14 b & 1.92 & 3.16 & 3.73 & 4.84 & 6.03 & 6.84 & 4.79 & 4.88 & 4.96 & 5.08 & 5.33 \\
HAT-P-37 b & 1.92 & 3.11 & 3.67 & 4.85 & 6.09 & 6.87 & 4.79 & 4.88 & 4.96 & 5.08 & 5.33 \\
HATS-32 b & 1.92 & 3.14 & 3.79 & 5.16 & 6.38 & 7.08 & 4.79 & 4.88 & 4.96 & 5.08 & 5.33 \\
CoRoT-12 b & 2.12 & 3.18 & 4.04 & 5.55 & 6.72 & 7.56 & 4.80 & 4.91 & 5.00 & 5.11 & 5.27 \\
CoRoT-29 b & 2.61 & 3.16 & 3.86 & 5.13 & 6.31 & 9.36 & 4.79 & 4.89 & 4.98 & 5.08 & 5.22 \\
HAT-P-20 b & 1.73 & 3.13 & 3.70 & 4.75 & 5.85 & 9.78 & 4.79 & 4.89 & 4.97 & 5.08 & 5.35 \\
WASP-23 b & 1.92 & 3.16 & 3.78 & 4.93 & 5.99 & 6.71 & 4.79 & 4.88 & 4.96 & 5.08 & 5.33 \\
HATS-13 b & 2.61 & 3.13 & 3.73 & 4.99 & 6.18 & 7.05 & 4.79 & 4.89 & 4.98 & 5.08 & 5.22 \\
HATS-28 b & 2.35 & 3.17 & 3.80 & 4.98 & 6.29 & 7.20 & 4.78 & 4.88 & 4.98 & 5.12 & 5.29 \\
Kepler-426 b & 1.92 & 3.13 & 3.72 & 5.29 & 6.54 & 7.36 & 4.79 & 4.88 & 4.96 & 5.08 & 5.33 \\
Kepler-44 b & 2.35 & 3.10 & 3.61 & 4.71 & 5.73 & 6.51 & 4.78 & 4.88 & 4.98 & 5.12 & 5.29 \\
HAT-P-28 b & 2.35 & 3.12 & 3.68 & 4.96 & 6.27 & 7.03 & 4.78 & 4.88 & 4.98 & 5.12 & 5.29 \\
WASP-89 b & 2.35 & 4.34 & 4.82 & 5.34 & 6.15 & 8.37 & 4.78 & 4.88 & 4.98 & 5.12 & 5.29 \\
WASP-6 b & 2.12 & 3.21 & 4.08 & 5.83 & 6.86 & 7.55 & 4.80 & 4.91 & 5.00 & 5.11 & 5.27 \\
WASP-124 b & 2.89 & 3.10 & 3.67 & 4.65 & 5.68 & 6.30 & 4.75 & 4.89 & 5.00 & 5.11 & 5.23 \\
HAT-P-63 b & 2.61 & 3.12 & 3.67 & 4.91 & 6.12 & 6.75 & 4.79 & 4.89 & 4.98 & 5.08 & 5.22 \\
TOI-3757 b & 1.92 & 3.17 & 4.07 & 5.38 & 6.64 & 7.34 & 4.79 & 4.88 & 4.96 & 5.08 & 5.33 \\
Kepler-428 b & 2.12 & 3.12 & 3.65 & 4.74 & 5.93 & 6.75 & 4.80 & 4.91 & 5.00 & 5.11 & 5.27 \\
WASP-22 b & 2.12 & 3.10 & 3.54 & 4.51 & 5.51 & 6.37 & 4.80 & 4.91 & 5.00 & 5.11 & 5.27 \\
CoRoT-27 b & 1.56 & 3.10 & 3.53 & 4.39 & 6.26 & 10.72 & 4.79 & 4.90 & 4.98 & 5.09 & 5.36 \\
HATS-8 b & 1.40 & 3.25 & 4.10 & 5.47 & 6.78 & 8.01 & 4.78 & 4.89 & 5.00 & 5.14 & 5.34 \\
HAT-P-55 b & 2.12 & 3.16 & 3.85 & 5.04 & 6.18 & 7.06 & 4.80 & 4.91 & 5.00 & 5.11 & 5.27 \\
HATS-11 b & 2.35 & 3.19 & 4.03 & 5.28 & 6.36 & 7.07 & 4.78 & 4.88 & 4.98 & 5.12 & 5.29 \\
CoRoT-23 b & 2.89 & 4.91 & 5.45 & 6.58 & 9.73 & 11.66 & 4.75 & 4.89 & 5.00 & 5.11 & 5.23 \\
HAT-P-25 b & 2.35 & 3.20 & 3.95 & 5.02 & 6.02 & 6.79 & 4.78 & 4.88 & 4.98 & 5.12 & 5.29 \\
TOI-905 b & 2.61 & 3.13 & 3.65 & 4.66 & 5.88 & 6.72 & 4.79 & 4.89 & 4.98 & 5.08 & 5.22 \\
Kepler-425 b & 1.92 & 3.16 & 3.85 & 5.21 & 6.65 & 7.46 & 4.79 & 4.88 & 4.96 & 5.08 & 5.33 \\
HAT-P-54 b & 2.61 & 3.13 & 3.63 & 4.66 & 5.63 & 6.42 & 4.79 & 4.89 & 4.98 & 5.08 & 5.22 \\
HATS-47 b & 2.89 & 3.15 & 3.89 & 5.12 & 6.23 & 6.92 & 4.75 & 4.89 & 5.00 & 5.11 & 5.23 \\
TOI-3629 b & 2.61 & 3.12 & 3.73 & 4.68 & 5.70 & 6.67 & 4.79 & 4.89 & 4.98 & 5.08 & 5.22 \\
TOI-1296 b & 2.61 & 3.14 & 3.71 & 4.84 & 5.83 & 6.76 & 4.79 & 4.89 & 4.98 & 5.08 & 5.22 \\
HAT-P-19 b & 1.92 & 3.16 & 3.93 & 5.29 & 6.41 & 7.12 & 4.79 & 4.88 & 4.96 & 5.08 & 5.33 \\
HAT-P-58 b & 2.89 & 3.18 & 3.91 & 5.03 & 6.13 & 6.91 & 4.75 & 4.89 & 5.00 & 5.11 & 5.23 \\
CoRoT-5 b & 2.35 & 3.24 & 4.16 & 5.46 & 6.77 & 7.44 & 4.78 & 4.88 & 4.98 & 5.12 & 5.29 \\
K2-30 b & 2.61 & 3.12 & 3.63 & 4.64 & 5.48 & 6.14 & 4.79 & 4.89 & 4.98 & 5.08 & 5.22 \\
HAT-P-59 b & 1.27 & 3.10 & 3.57 & 4.72 & 5.96 & 10.67 & 4.75 & 4.90 & 5.01 & 5.13 & 5.31 \\
HAT-P-51 b & 2.61 & 3.25 & 4.06 & 5.46 & 6.50 & 7.33 & 4.79 & 4.89 & 4.98 & 5.08 & 5.22 \\
HATS-25 b & 2.35 & 3.13 & 3.85 & 4.96 & 6.24 & 7.10 & 4.78 & 4.88 & 4.98 & 5.12 & 5.29 \\
HATS-43 b & 2.12 & 3.16 & 3.87 & 5.28 & 6.45 & 7.26 & 4.80 & 4.91 & 5.00 & 5.11 & 5.27 \\
Kepler-12 b & 3.56 & 3.16 & 3.76 & 4.83 & 5.85 & 6.42 & 4.68 & 4.85 & 4.99 & 5.12 & 5.29 \\
WASP-151 b & 3.56 & 3.14 & 3.58 & 4.70 & 5.66 & 6.33 & 4.68 & 4.85 & 4.99 & 5.12 & 5.29 \\
HATS-29 b & 2.35 & 3.15 & 3.88 & 4.96 & 5.98 & 6.82 & 4.78 & 4.88 & 4.98 & 5.12 & 5.29 \\
HATS-5 b & 2.35 & 3.10 & 3.53 & 4.37 & 5.34 & 6.32 & 4.78 & 4.88 & 4.98 & 5.12 & 5.29 \\
CoRoT-16 b & 2.61 & 3.29 & 5.02 & 6.23 & 7.01 & 7.98 & 4.79 & 4.89 & 4.98 & 5.08 & 5.22 \\
TOI-2669 b & 3.21 & 3.14 & 3.86 & 5.00 & 6.16 & 6.96 & 4.71 & 4.87 & 5.02 & 5.15 & 5.28 \\
K2-140 b & 2.61 & 3.10 & 3.66 & 4.72 & 5.59 & 6.35 & 4.79 & 4.89 & 4.98 & 5.08 & 5.22 \\
TOI-559 b & 5.38 & 3.60 & 4.02 & 4.76 & nan & nan & 4.72 & 4.89 & 5.08 & 5.29 & 5.43 \\
WASP-59 b & 0.50 & 3.05 & 3.33 & 4.14 & nan & nan & 5.03 & 5.38 & 5.75 & 6.39 & 7.84 \\
TOI-1268 b & 0.50 & 3.07 & 3.34 & 3.89 & nan & nan & 5.03 & 5.38 & 5.75 & 6.39 & 7.84 \\
TOI-2158 b & 0.50 & 3.10 & 3.49 & 4.29 & nan & nan & 5.03 & 5.38 & 5.75 & 6.39 & 7.84 \\
CoRoT-30 b & 0.50 & 3.08 & 3.38 & 4.24 & nan & nan & 5.03 & 5.38 & 5.75 & 6.39 & 7.84 \\
K2-132 b & 0.50 & 3.04 & 3.27 & 3.90 & nan & nan & 5.03 & 5.38 & 5.75 & 6.39 & 7.84 \\
WASP-185 b & 10.00 & 4.33 & 4.82 & 5.33 & nan & nan & 4.91 & 5.02 & 5.13 & 5.71 & 6.71 \\
TOI-172 b & 2.89 & 3.08 & 3.32 & 3.80 & nan & nan & 4.75 & 4.89 & 5.00 & 5.11 & 5.23 \\
\end{tabular}

        }
    }
\end{table*}

The above result ignores the thermal evolution of the planet. Tidal heating of a
planet can affect its size and hence change the rate of orbital evolution
\citep{alvarado2019orbital, rozner2022inflated}. If the planets are initially
large, the cooling process tries to shrink their radius. Planets start out large
and cool and shrink over time. Tidal heating can slow down this shrinking or
even reverse it.  As a result, planets would have had larger radii in the past
than they do today, potentially lasting long enough to imply noticeably faster
circularization for a given value of $Q_{pl}'$. The extent to which this would
affect the inferred value of $Q_{pl}'$ depends on how long these larger radii
persisted and on the physical mechanisms driving tidal dissipation. For example,
if visco-elastic dissipation in the solid core of planets dominates, tidal
heating will have relatively little effect on the circularization rate, i.e. the
effective $Q_{pl}'$ will scale with the planet's radius in such a way as to
approximately cancel the radius dependence from the rate of tidal
circularization (assuming the core mass and radius do not change as the planet
cools). If inertial mode dissipation dominates, it also depends on the ratio of
the core's radius to the whole radius of the planet
\citep{guenel2014unravelling}, though not sufficiently as to completely cancel
the radius dependence.

We have assumed that the observational uncertainties of the
physical properties of the systems (planet and star mass, age, metallicity,
planet radius, orbital eccentricity) are independent Gaussian distributions. By
ignoring the correlations between planetary and stellar properties we are
inflating the uncertainty in the measurement of the tidal quality factor. For
example, tidal evolution is only sensitive to the planet-to-star mass ratio,
which is usually constrained much more tightly by radial velocity observations than
each mass separately. Sampling each mass independently results in a much
broader distribution of mass ratios. Similarly, for many systems, the planet-to-star radius ratio is not included in the Exoplanet Archive, so we impose a
Gaussian prior on the planet radius directly. Once again this results in a broader
distribution of planet-to-star size ratios, translating to larger inferred
uncertainty in $\log_{10}Q_{pl}'$. For those systems for which the square radius
ratio is provided, we impose a prior on it and infer the planet radius from the
stellar radius the built-in POET stellar evolution predicts for the sampled
stellar mass, age, and metallicity.

\section{Comparison to Prior Tidal Dissipation Constraints}
\label{sec:comparison}

\begin{figure}
    \includegraphics[width=\columnwidth]{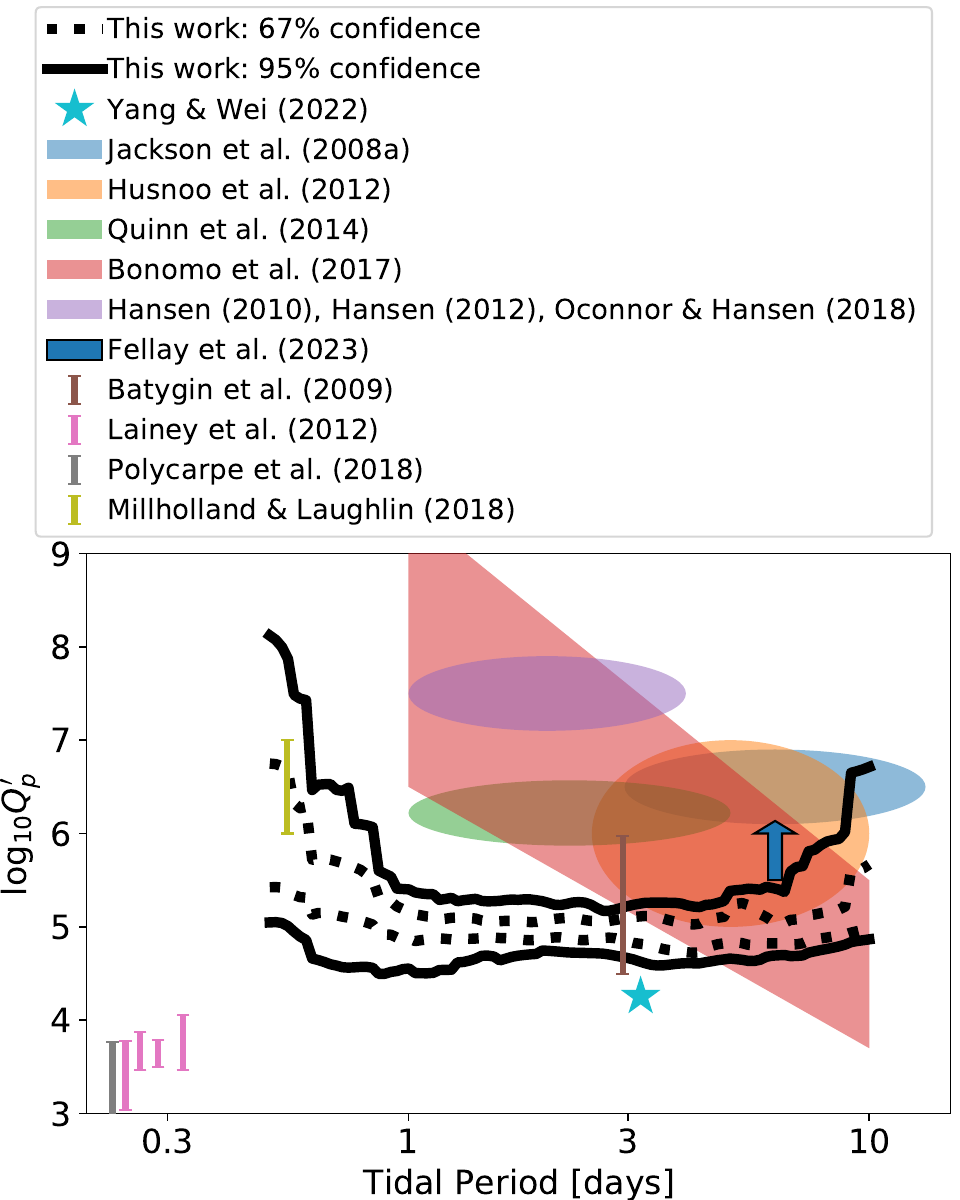}
    \caption{
        The combined constraint on $\log_{10} Q'_{pl}$ as a function of the
        tidal period from our analysis of 77 HJ systems (see text) demonstrating
        the expected precision and period range over which tidal dissipation
        will be constrained by this project. The solid lines show the 95\%
        confidence interval (2.3\% and 97.7\% percentiles of the combined
        posterior distribution), and the dotted lines show the 68\% confidence
        interval (15.9\% and 84.1\% percentiles).
    }
    \label{fig:comparison}
\end{figure}

Figure \ref{fig:comparison} summarizes the comparison between our constraints
and the constraints on $Q_{pl}'$ found in the literature. Below we discuss in
more detail each of the constraints shown in the figure as well as two
additional analyses which only provide a lower limit to $Q_{pl}'$.

\citet{jackson2008tidal} tuned the values of $Q_{pl}'$ and $Q_{\star}'$ (assumed
constant) by requiring that the initial eccentricity distribution for orbital
semi-major axes below 0.2 AU matches that of larger orbits. They found the best
fit based on a Kolmogorov-Smirnov test for $Q_{pl}'=10^{6.5}$ and
$Q_{\star}'=10^{5.5}$. Their estimated value of $Q_{pl}$ is an order of
magnitude larger than our finding. The majority of the planets used by
\citet{jackson2008tidal} were detected through radial velocity observations. As
a result, for the majority of their planets, only a lower limit was available
for the planetary mass due to the unknown orbital inclination, and the planetary
radius was not directly constrained by observations. For such planets,
\citet{jackson2008tidal} assumed the planet mass is equal to the minimum mass,
and for gas giant planets assumed a radius of 1.2 Jupiter radii. Perhaps
less importantly, the tidal evolution model used by \citet{jackson2008tidal}
ignored stellar evolution and only included first-order eccentricity terms.
Since these authors are investigating the same tidal signature as us, the
difference in inferred $Q_{pl}'$ values must be driven by some combination of
the superior parameters provided by transiting planets, the larger sample size
at our disposal, the somewhat different statistical treatment, and to a
lesser extent the improved tidal model.

\citet{Hansen_10, Hansen_12, Oconnor_Hansen_18} used a different
parameterization for tides than $Q_{pl}'$, based on
\citet{eggleton1998equilibrium}, equivalent to a constant time lag, i.e.
$Q_{pl}' \propto P_{tide}$. They calibrated their tidal model based on matching
the period-eccentricity envelope for a fiducial star-planet combination to the
observed population of gas giants and then refined the constraints by modeling
individual objects in detail. Converting their constraints to present-day
$Q_{pl}'$, these authors find $7 < \log_{10}Q_{pl}' < 8$, which is between two
and three orders of magnitude larger (less dissipation) than our results.
However, the strictest lower limits in there are consistent with our constraints.

\citet{husnoo2012observational} did not directly constrain the tidal quality
factor. They showed observations are roughly consistent with $\log_{10} Q_{pl}'
\sim 6$ and $\log_{10} Q_\star' \sim 4$. Since they did not explore what range of
values would still be consistent with observations, it is difficult to tell if
and how much tension there is between our constraints on $\log_{10} Q'_{pl}$ and
theirs.

\cite{quinn2014hd} argue that the statistical difference in the distribution of
eccentricities, when comparing systems younger and older than their
corresponding circularization timescale, is maximized by $\log_{10} Q_{pl}'
=6.14^{+0.41}_{-0.25}$.  That is approximately one order of magnitude higher
than the value of $Q_{pl}$ determined by us.

\citet{bonomo2017gaps} re-derived eccentricities of 231 transiting gas giant
exoplanets. Under the assumption $Q_{pl}'=const.$, for systems with clearly
circular orbits, they obtained upper limits for $Q_{pl}'$ by requiring that the
tidal circularization timescales should be shorter than the age of the system;
for systems with clearly eccentric orbits, lower limits to $Q_{pl}'$ follow by
requiring that the circularization timescales exceed the age of the system. The
resulting upper and lower limits roughly corresponding to the red quadrangle are
shown in Fig.~\ref{fig:comparison}. Their results hint at a frequency
dependence, though given the simple timescale arguments used and the fact that
only age uncertainties were accounted for in the constraints, which may simply
be an artifact. Regardless, our constraints overlap with those of
\citet{bonomo2017gaps} for tidal periods of a few days, but for the shortest
period systems we seem to find significantly more dissipation than their
analysis suggests.

\citet{batygin2009determination} investigated the internal structure of a
transiting planet (HAT-P-13 b) perturbed by another planet. With a mass of $0.85
M_{Jup}$, a radius of $1.28 R_{Jup}$, and orbital period of 2.91 days, HAT-P-13
b falls squarely within the class of planets we investigate here. They argued
that due to the perturbation by the other planet, the aphelion-perihelion
joining lines of both planets' orbits get aligned, forcing the orbits of both
planets to precess at the same rate. The apsidal precession rate significantly
depends on general relativistic effects, as well as the effects of the
gravitational quadrupole fields created by the transiting planet’s tidal and
rotational distortions. That can be a probe into the internal structure of the
transiting planet. The authors concluded that the tidal quality factor of the
transiting planet must satisfy $4 < \log_{10}Q_{pl} < 5.5$, which combined
with the corresponding values for the tidal Love number
\citet{batygin2009determination} find implies $4.5 < \log_{10}Q_{pl}' < 6$,
fully consistent with our constraints.

\citet{lainey2012strong} claim that recent high-precision measurements of the
positions of Saturn's moons show their orbits are evolving. Attributing that
evolution to tidal dissipation within Saturn, they find a tidal quality factor for the tidal frequency of each moon separately. Assuming the same value applies to all moons
they conclude that for Saturn, $1/Q'=(2.3 \pm 0.7)\times10^{-4}$. This
constraint is at significantly higher tidal frequencies than we probe here, so it is
not directly comparable, but the inferred dissipation is about one order of
magnitude larger (lower $Q_{pl}'$) than a naive extrapolation of our results
would indicate.

\citet{millholland2018obliquity} suggest the observed rapid inspiral of WASP-12 b could be due to obliquity tides if the obliquity of the planet is maintained
by an interaction with an undetected additional planet in the system. Assuming
this interpretation is correct, they show that the required tidal quality factor
for the planet is $Q' \thicksim 10^{6}-10^{7}$. The relevant tidal period for
WASP-12 b is below the range where our analysis produces tight constraints and is
formally well within the broad range of $Q_{pl}'$ values our analysis gives at
these tidal frequencies.

\citet{yang2022transit} (not shown in Fig.~\ref{fig:comparison}) argued that the
orbital period decay of XO-3 b, which is evident by its Transit Timing Variation
(TTV), can be explained by tidal interactions. If tidal interaction is the
mechanism behind this observed TTV, and if orbital period decay is due to the tide
in the planet, then the $Q'_{pl}$ of the planet
would be $1.8\times10^{4}\pm8\times10^{2}$. On the other extreme, if the tides
in the star are solely responsible, then
$Q_\star'=1.5\times10^{5}\pm6\times10^{3}$.  Since the TTVs can be due to a
combination of planetary and stellar tides, this result can be interpreted as
$Q_{pl}' \geq 1.8\times10^{4}\pm8\times10^{2}$.  Our constraints would require
a significant contribution from stellar tides to explain the observed TTVs.

\citet{fellay2023constraints} (not shown in Fig.~\ref{fig:comparison}) carried
out a very similar analysis to the one used here to find a lower limit on $Q_{pl}'$
for a single exoplanet, Kepler-91 b, with the benefit of detailed
asteroseismic characterization of the host star Kepler-91. These authors found
that in order for the observed eccentricity of Kepler-91 b to survive to the
present day $Q_{pl}'$ must exceed $4.5^{+5.8}_{-1.5}\times10^{5}$. Our combined
constraints clearly satisfy this limit.

\section{Conclusions}
\label{sec:conclusions}

In this project, we constrained the tidal dissipation quality factor of hot
Jupiter exoplanets. We selected a sample of exoplanet systems for which the
planet is clearly a gas giant with a short orbital period and the parent star is
a main sequence star with a radiative core and convective spherical shell. We
proposed an empirical model for frequency-dependent tidal dissipation and
carried out a detailed system-by-system Bayesian analysis to account for all
observational uncertainties. We extracted individual constraints. Each
individual constraint is not very informative; however, since the planets in our
sample were all chosen to be very similar, it is reasonable to require that
their dissipation will also be similar. We, therefore, combined the individual
constraints, requiring a unified tidal dissipation prescription. We found no
clear sign of the dependence of the tidal quality factor on tidal frequency. For
the range of tidal period from 0.8 to 7 days, the modified tidal quality factor
for the planets belonging to the systems is found to be $\log_{10}Q_{pl}'=5.0
\pm 0.5$. The combined constraint was found to be consistent with each
individual constraint; it is thus capable of explaining the observed
eccentricity envelope while simultaneously allowing the observed eccentricity of
each system to survive to the present day.

\section*{Acknowledgements}

This research was supported by NASA grant 80NSSC18K1009.

The authors acknowledge the Texas Advanced Computing Center (TACC)\footnote{\url{http://www.tacc.utexas.edu}} at The
University of Texas at Austin for providing HPC resources that have contributed
to the research results reported within this paper.

This research has made use of the NASA Exoplanet Archive, which is operated by the California Institute of Technology, under contract with the National Aeronautics and Space Administration under the Exoplanet Exploration Program.

\section*{Data Availability}

We have created a zenodo archive \citep{zenodo_results} to accompany this
article, which provides machine-readable tables of the following:

\begin{itemize}
    \item The generated MCMC samples for each exoplanet system in the original
        HDF5 format produced by the emcee package \citep{foreman2013emcee}
        (see \url{https://emcee.readthedocs.io/en/stable/user/backends/}).
    \item The 2.3\%, 15.9\%, 84.1\%, and 97.7\% quantiles of $\log_{10}Q_{pl}'$
        for the given binary as a function of the tidal period for each binary (i.e.
        the coordinates defining the quantile curves in
    Fig.~\ref{fig:indivCons1} and \ref{fig:indivCons2}).
    \item The burn-in period for each quantile vs. tidal period (i.e. the
        coordinates defining the burn-in curves in Fig.~\ref{fig:burn1} and
    \ref{fig:burn2}).
    \item The estimated standard deviation of the fraction of samples below each
        of the 4 target quantiles for each binary (i.e. the coordinates defining
        the curves in Fig.~\ref{fig:sigmacdf1} and \ref{fig:sigmacdf2}).
    \item The 2.3\%, 15.9\%, 84.1\%, and 97.7\% percentiles of the combined
        $\log_{10}Q_{pl}'$ constraint from all binaries vs tidal period (i.e. the
        coordinates of the quantile curves in Fig.~\ref{fig:comb}).
    \item The combined constraint 2.3\%, 15.9\%, 84.1\%, and 97.7\% percentiles
        and the 2.3\% and 97.7\% percentiles of individual constraints at the
        tidal periods where the difference between the latter is smallest (i.e.
            the coordinates defining the curves and points in
        Fig.~\ref{fig:indivVsComb}).
\end{itemize}

The version of POET used for calculating the orbital evolution is available
through zenodo archive \citet{zenodo_poet}.




\bibliographystyle{mnras}
\bibliography{references} 







\bsp	
\label{lastpage}
\end{document}